\documentclass[twocolumn]{aastex631}

\usepackage{amsmath,amstext}
\usepackage[T1]{fontenc}
\usepackage[figure,figure*]{hypcap}
\usepackage{xspace}
\usepackage{xcolor}
\usepackage{hyperref}
\usepackage{float}
\usepackage{wrapfig} 
\usepackage{longtable}
\usepackage{soul} 

\newcommand\aastex{AAS\TeX}

\newcommand{\rphk}{log($\mathrm{R}'_\mathrm{HK}$)\xspace}
\newcommand{\logg}{log($g$)\xspace}
\newcommand{\teff}{$\mathrm{T}_\mathrm{eff}$\xspace}
\newcommand{\feh}{[Fe/H]\xspace}

\newcommand{\cahk}{Ca~II~H\,\&\,K\xspace}
\newcommand{\rstar}{\ensuremath{R_\star}\xspace}
\newcommand{\mstar}{\ensuremath{M_\star}\xspace}
\newcommand{\rsun}{\ensuremath{R_\sun}\xspace}
\newcommand{\msun}{\ensuremath{M_\sun}\xspace}

\begin{document}

\title{Template \aastex Article with Examples: 
v6.3.1\footnote{Released on March, 1st, 2021}}

\newcommand{\USQ}{Centre for Astrophysics, University of Southern Queensland, Toowoomba, QLD, Australia}
\newcommand{\CIT}{California Institute of Technology, Pasadena, CA 91125, USA}

\title{The California Legacy Survey. V. Chromospheric Activity Cycles in Main Sequence Stars}

\author[0000-0002-0531-1073]{Howard Isaacson}
\affiliation{501 Campbell Hall, University of California at Berkeley, Berkeley, CA 94720, USA}
\affiliation{\USQ}

\author[0000-0001-8638-0320]{Andrew W. Howard}
\affiliation{\CIT}

\author[0000-0003-3504-5316]{Benjamin~Fulton}
\affiliation{NASA Exoplanet Science Institute/Caltech-IPAC, MC 314-6, 1200 E. California Blvd., Pasadena, CA 91125, USA}

\author[0000-0003-0967-2893]{Erik~A.~Petigura}
\affiliation{Department of Physics \& Astronomy, University of California Los Angeles, Los Angeles, CA 90095, USA}

\author[0000-0002-3725-3058]{Lauren M. Weiss}
\affiliation{Department of Physics and Astronomy, University of Notre Dame, Notre Dame, IN 46556, USA}

\author[0000-0002-7084-0529]{Stephen R. Kane}
\affiliation{Department of Earth and Planetary Sciences, University of California, Riverside, CA 92521, USA}
\affiliation{\USQ}

\author[0000-0002-7084-0529]{Brad Carter}
\affiliation{\USQ}

\author[0000-0001-7708-2364]{Corey~Beard}
\altaffiliation{NASA FINESST Fellow}
\affiliation{Department of Physics \& Astronomy, University of California Irvine, Irvine, CA 92697, USA}

\author[0000-0002-8965-3969]{Steven~Giacalone}
\affiliation{Department of Astronomy, California Institute of Technology, Pasadena, CA 91125, USA}
\altaffiliation{NSF Astronomy and Astrophysics Postdoctoral Fellow}

\author[0000-0002-4290-6826]{Judah Van Zandt}
\affiliation{Department of Physics \& Astronomy, University of California Los Angeles, Los Angeles, CA 90095, USA}

\author[0000-0001-8898-8284]{Joseph~M.~Akana~Murphy}
\affiliation{Department of Astronomy and Astrophysics, University of California, Santa Cruz, CA 95064, USA}
\altaffiliation{NSF Graduate Research Fellow}

\author[0000-0002-8958-0683]{Fei~Dai} 
\affiliation{Division of Geological and Planetary Sciences,
1200 E California Blvd, Pasadena, CA, 91125, USA}
\affiliation{Department of Astronomy, California Institute of Technology, Pasadena, CA 91125, USA}
\altaffiliation{NASA Sagan Fellow}

\author[0000-0003-1125-2564]{Ashley~Chontos}
\affiliation{Institute for Astronomy, University of Hawai`i, 2680 Woodlawn Drive, Honolulu, HI 96822, USA}
\affiliation{Department of Astrophysical Sciences, Princeton University, 4 Ivy Lane, Princeton, NJ 08544, USA}
\altaffiliation{Henry Norris Russell Fellow}

\author[0000-0001-7047-8681]{Alex~S.~Polanski}
\affil{Department of Physics and Astronomy, University of Kansas, Lawrence, KS 66045, USA}

\author[0000-0002-7670-670X]{Malena~Rice}
\affil{Department of Astronomy, Yale University, New Haven, CT 06511, USA}

\author[0000-0001-8342-7736]{Jack Lubin}
\affiliation{Department of Physics \& Astronomy, University of California Los Angeles, Los Angeles, CA 90095, USA}
\affiliation{Department of Physics \& Astronomy, The University of California Irvine, Irvine, CA 92697, USA}

\author[0000-0002-4480-310X]{Casey Brinkman}
\affiliation{Department of Astronomy, 501 Campbell Hall, University of California, Berkeley, CA 94720, USA}

\author[0000-0003-3856-3143]{Ryan~A.~Rubenzahl}
\altaffiliation{NSF Graduate Research Fellow}
\affiliation{Department of Astronomy, California Institute of Technology, Pasadena, CA 91125, USA}

\author[0000-0002-3199-2888]{Sarah Blunt}
\affiliation{Center for Interdisciplinary Exploration and Research in Astrophysics (CIERA) and Department of Physics and Astronomy, Northwestern University, Evanston, IL 60208, USA}

\author[0000-0001-7961-3907]{Samuel W. Yee}
\affiliation{Department of Astrophysical Sciences, Princeton University, 4 Ivy Lane, Princeton, NJ 08544, USA}

\author[0000-0003-2562-9043]{Mason~G.~MacDougall}
\affiliation{Department of Physics \& Astronomy, University of California Los Angeles, Los Angeles, CA 90095, USA}

\author[0000-0002-4297-5506]{Paul~A.~Dalba}
\affiliation{Department of Astronomy and Astrophysics, University of California, Santa Cruz, CA 95064, USA}

\author[0000-0003-0298-4667]{Dakotah Tyler}
\affiliation{Department of Physics and Astronomy, University of California, Los Angeles, CA 90095, USA}

\author[0000-0002-4480-310X]{Aida Behmard}
\affiliation{Department of Astrophysics, American Museum of Natural History, 200 Central Park West, Manhattan, NY 10024, USA}

\author[0000-0002-9751-2664]{Isabel Angelo}
\affiliation{Department of Physics \& Astronomy, University of California Los Angeles, Los Angeles, CA 90095, USA}

\author[0000-0001-9771-7953]{Daria~Pidhorodetska} 
\affiliation{Department of Earth and Planetary Sciences, University of California, Riverside, CA 92521, USA}
\affiliation{NASA FINESST Fellow}

\author[0000-0002-7216-2135]{Andrew W. Mayo}
\affiliation{Astronomy Department, University of California, Berkeley, CA 94720, USA}
\affiliation{Centre for Star and Planet Formation, Natural History Museum of Denmark \& Niels Bohr Institute, University of Copenhagen, Øster Voldgade 5-7, DK-1350 Copenhagen K., Denmark}

\author[0000-0002-5034-9476]{Rae~Holcomb}
\affiliation{Department of Physics \& Astronomy, University of California Irvine, Irvine, CA 92697, USA}

\author[0000-0002-1845-2617]{Emma~V.~Turtelboom}
\affiliation{Department of Astronomy, 501 Campbell Hall, University of California, Berkeley, CA 94720, USA}

\author[0000-0002-0139-4756]{Michelle~L.~Hill}
\affiliation{Department of Earth and Planetary Sciences, University of California, Riverside, CA 92521, USA}

\author[0000-0002-0514-5538]{Luke G. Bouma}
\affiliation{Department of Astronomy, MC 249-17, California Institute of Technology, Pasadena, CA 91125, USA}

\author[0000-0002-2696-2406]{Jingwen,Zhang}
\altaffiliation{NASA FINESST Fellow}
\affiliation{Institute for Astronomy, University of Hawai’i, 2680 Woodlawn Drive, Honolulu, HI 96822, USA}

\author{Ian~J.~M.~Crossfield}
\affiliation{Department of Physics \& Astronomy, University of Kansas, 1082 Malott, 1251 Wescoe Hall Dr., Lawrence, KS 66045, USA}


\author[0000-0003-2657-3889]{Nicholas~Saunders}
\affiliation{Institute for Astronomy, University of Hawai`i, 2680 Woodlawn Drive, Honolulu, HI 96822, USA}
\altaffiliation{NSF Graduate Research Fellow}







\correspondingauthor{Howard Isaacson}
\email{hisaacson@berkeley.edu}

\begin{abstract}

We present optical spectroscopy of 710 solar neighborhood stars collected over twenty years to catalog chromospheric activity and search for stellar activity cycles. 
The California Legacy Survey stars are amenable to exoplanet detection using precise radial velocities, and we present their Ca II H\&K time series as a proxy for stellar and chromospheric activity. Using the HIRES spectrometer at Keck Observatory, we measured stellar flux in the cores of the Ca II H\,\&\,K lines to determine $S$-values on the Mt.\ Wilson scale and the \rphk metric, which is comparable across a wide range of spectral types. From the 710 stars, with 52372 observations, 285 stars are sufficiently sampled to search for stellar activity cycles with periods of 2--25 years, and 138 stars show stellar cycles of varying length and amplitude. $S$-values can be used to mitigate stellar activity in the detection and characterization of exoplanets. We use them to probe stellar dynamos and to place the Sun’s magnetic activity into context among solar neighborhood stars. Using precise stellar parameters and time-averaged activity measurements, we find tightly constrained cycle periods as a function of stellar temperature between \rphk of -4.7 and -4.9, a range of activity in which nearly every star has a periodic cycle. These observations present the largest sample of spectroscopically determined stellar activity cycles to date.

\end{abstract}

\keywords{Chromospheric Activity, Stellar Activity Cycles, Stellar Astrophysics}

\section{Introduction}
\label{sec:intro}

Long-term, ground-based spectroscopic surveys are a pathway to understanding the Sun and its planets in the context of the solar neighborhood and to finding Earth-analog exoplanet systems.
Such surveys can probe the depth of the convective zone, detect differential rotation, and track sun-like stellar activity cycles. Chromospheric activity studies provide fascinating insights into the subsurface layers of stars that are not directly observable.  Over the last two decades, these studies have been buoyed by radial velocity (RV) searches for exoplanets due to the collection of time cadence observations that include spectral information that can be used for both measuring precise RVs of stars and monitoring stellar chromospheric activity (\citealt{Rosenthal2021}, hereafter CLS1; \citealt{GomesdaSilva2021}). 

Nightly surveying of the chromospheric activity of nearby stars in the Mt.\ Wilson H\,\&\,K project began in 1966 \citep{Wilson1968} and continued for several decades \citep{Vaughan1978}.  This survey detected variable stellar lines and identified the link between the \cahk lines and the solar chromosphere \citep{Eberhard1913}. After decades of data collection on F2--M2 type stars -- an effort necessary to identify stellar activity cycles in some G0--K5 stars with activity periods similar to the Sun's eleven-year solar cycle -- the results were summarized \citep{Duncan1991} and the first catalog of stellar activity cycles was published \citep{Baliunas1995}. Out of 111 solar-type stars searched, ``...52 showed cycles, 31 are flat or have linear trends.''  Another 29 stars had non-periodic, variable activity. The conclusions put the Sun's activity cycle into the broader perspective of sun-like stars in the solar neighborhood, showing that stellar activity cycles are common. 

Several long-term ground-based surveys have contributed to our understanding of stellar magnetic activity. Identification of stellar activity cycles using Mt.\ Wilson data combined with California Planet Search (CPS) data from the High Resolution Echelle Spectrograph (HIRES) at the W.\,M.\ Keck Observatory yielded baselines of 50 years for 59 stars \citep{Baum2022}. Time-series spectroscopic observations of sun-like stars include a survey of 800 southern solar-type stars within 50 pc \citep{Henry1996} and 143 sun-like stars from 1996-2007 \citep{Hall2007}. These studies focused on measuring average stellar variability, not stellar cycles.  Fifty-three previously identified activity cycles were analyzed using S-values from the Mt. Wilson H\&K Project and the HARPS telescope \citep{BoroSaikia2018}, but even the extended HARPS baseline was insufficient for identifying stellar new activity cycles that span years to decades. See \cite{Jeffers2023} for a comprehensive review of stellar activity cycles.

Stellar activity and planet searches that use the RV technique have contributed to our knowledge of Jupiter-mass planets with orbital periods of more than 10 years and to the identification of solar-like stellar cycles \citep{Wright2008,Fulton2021}. \cite{Wright2004} and \cite{Isaacson2010} presented activity catalogs from Keck/HIRES and began to quantify the relationship between RV jitter and chromospheric activity. \cite{Luhn2020} examined 600 California Planet Search stars to make RV jitter assessments that included dependence on stellar surface gravity.  Summary of the ground-based spectroscopic survey of the AMBRE-HARPS sample \cite{GomesdaSilva2021} resulted in an activity catalog of planet search stars in the southern hemisphere, with stellar activity time-series analysis forthcoming. Detecting Jupiter analogs requires forward thinking surveys and understanding their dynamical impact in multi-planet systems will inform the study of solar-like planetary systems \citep{Kane2023}.

In an analysis of southern hemisphere planet search stars similar to the northern hemisphere sample presented here \cite{Lovis2011} analyzed seven years of High Accuracy Radial velocity Planet Searcher (HARPS) S-values for 304 FGK type stars and presented a catalog of 99 magnetic cycles and analysis of the stellar activity impact on precise RVs. Using the H-alpha line as an activity metric \citep{Robertson2013} searched 93 K and M type stars using the High Resolution Spectrograph on the Hobby-Eberly Telescope at McDonald Observatory and identified examples of how activity cycles can mimic giant planets. These two catalogs provide examples of how planet search data has been used to study magnetic activity.

Only with long-term baselines of activity and RVs are the periodic signals of planets distinguished from quasiperiodic activity signals. In some cases, a stellar activity cycle is correlated with the RVs, making the planet interpretation ambiguous \citep{Rosenthal2021}. \cite{Kane2016} identified a stellar activity cycle in HD 99492, a planet-hosting system, while \cite{Dragomir2012} found a photometric activity cycle. Correlations between RVs and S-values over a single period of the planet's orbit or the stellar activity cycle are difficult to disentangle \citep{Wright2008,Fulton2015}. But, if the baseline is extended sufficiently, the activity cycle may go out of phase, while a planet will maintain strict periodicity \citep{Wright2015}. Stellar activity cycles have been probed by other spectral features that are sensitive to activity. The M-dwarf GJ 328 has a confirmed planet along with a stellar activity cycle that was identified with H-alpha line measurements. The CARMENES planet search, which focuses on M dwarfs, produced a catalog of \rphk measurements to assist in the interpretation of planet candidates \citep{Perdelwitz2021}.

Wide-field, space-based photometry is now available to search for transiting planets, measure stellar rotation periods, and monitor stellar activity. Such photometry is particularly useful for determination of stellar rotation periods \citep{McQuillan2014,Angus2018}, and has revolutionized rotation studies.  The Kepler data can be searched for stellar activity cycles, with thousands of cycle candidates \citep{Shen2022}, but the four-year duration of the Kepler mission makes it difficult to find solar-like cycles. 

Long-term ground-based photometry can be used to find stellar activity cycles, but is only sensitive to cycles for stars with large spot coverage such as M-dwarfs.  \cite{Irving2023} examined activity cycles for a collection of M dwarf stars. The coolest M-dwarfs, M4 and later, require a different mechanism for magnetic field generation than solar-type stars since they lack a radiative-convective boundary. Spectral analyses of the \cahk lines probe the chromosphere and the magnetic activity below the observable stellar surface, which is complementary to the results of photometric surveys. Photometric studies of activity cycles in M-dwarfs have identified cycles in fully convective M-dwarfs with masses as low as 0.12 \msun \citep{Savanov2012,suarezmascareno2016,suarezmascareno2018,Wargelin2017}.

Ground-based photometric surveys were used to calibrate the age-activity-rotation relationship \citep{Barnes2007,Mamajek2008} by correlating stellar rotation periods of open-clusters with well-determined ages.  Time-averaged chromospheric flux measurements have been used to parameterize the physical mechanisms at work below the observable stellar photosphere. The Rossby number, the ratio of the rotation period to the convective turnover time \citep{Noyes1984}, is the standard metric for quantifying magnetic activity and its relationship to stellar rotation.  

Observations of stellar rotation periods, combined with stellar activity cycles, are fleshing out the magnetic activity evolution of stars as their rotation periods decline over time through weakened magnetic braking \citep{VanSaders2016,Metcalfe2022,David2022}. By combining \cahk measurements with rotation periods, and direct measurements of the magnetic field through spectropolarimetry \citep{Marsden2014,Metcalfe2024}, more complete explanations of the stellar dynamo are now coming into focus.  We add to the observational evidence that can be used to understand main-sequence magnetic changes and potentially to explaining weakened magnetic breaking.

We present twenty years of stellar chromospheric activity time-series for 710 nearby (median distance of 30 pc), main sequence FGKM stars, analyze average activity in terms of fundamental stellar properties and search for activity cycles like the Sun's 11-year cycle. Section \ref{sec:observations} discusses the observations and data quality. Section \ref{sec:activity_sample} discusses the CLS1 stellar sample and compare it to previously published works. Section \ref{sec:cycles_sample} discusses our 285 star sample that is searched for cycles. Section \ref{sec:starswithcycles} explores the activity cycles in terms of the stellar properties for 138 stars with detected cycles and Section \ref{sec:discussion} reveals the relationship between, cycle period and \teff.

\section{Observations}      
\label{sec:observations}

\begin{figure} 
\includegraphics[width = 0.9\columnwidth]{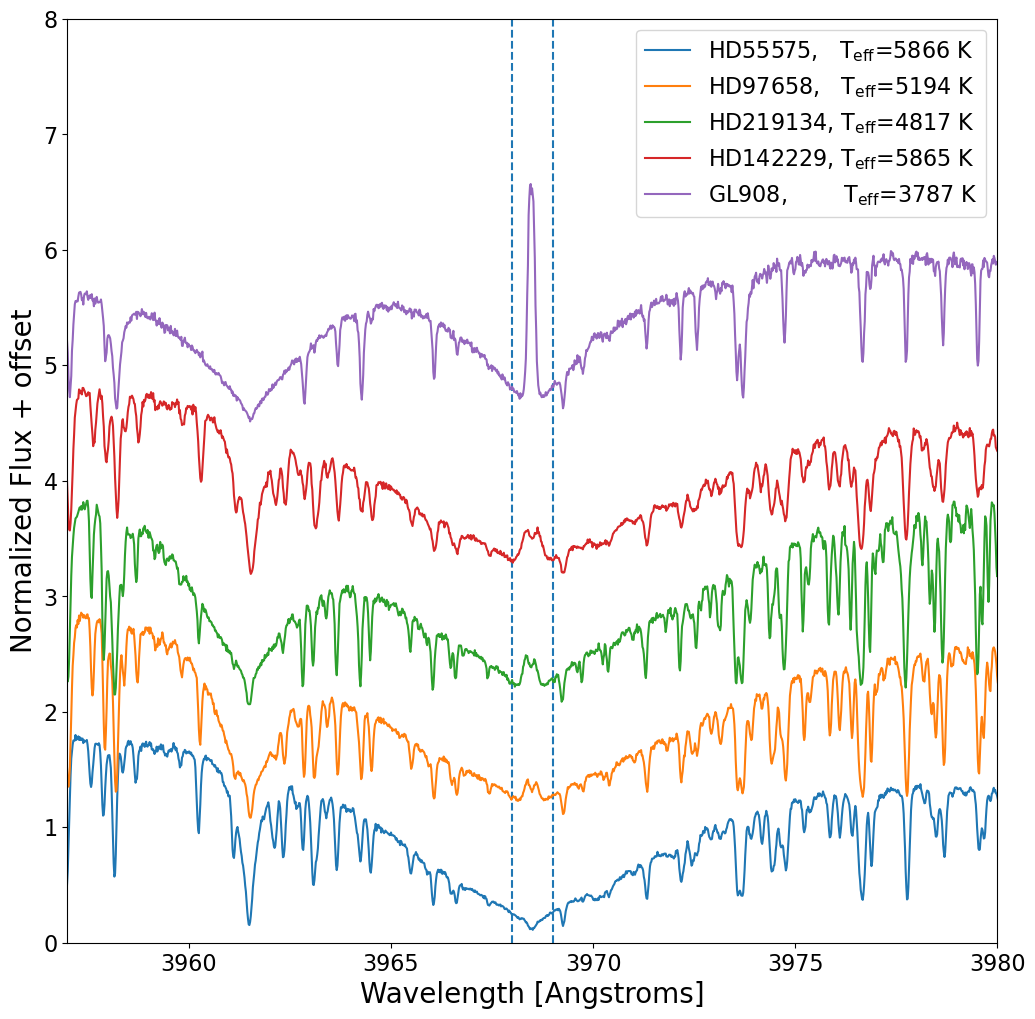}
\caption{ 
Ca H-lines for a variety of effective temperatures from our survey, offset vertically for clarity. The dotted lines mark the center of the extracted flux region for the H-line. From bottom to the top are HD 55575 (\teff = 5866 K, S-value = 0.156), HD 97658 (\teff = 5194 K, S-value = 0.186), HD 219134 (\teff = 4817 K , S-value = 0.246 ), HD 142229 (\teff = 5865 K, S-value = 0.364), and GL 908 (\teff = 3787 K, S-value = 0.54).
}
\label{fig:hline_profile} 
\end{figure}

\subsection{Data Source and Quality}
\label{sec:data_quality}

The CLS1 paper provided RVs measured from data collected from the middle of three detectors (4976--6421 \AA), and the S-values were simultaneously measured using data from the blue detector (3642--4797 \AA ). We used an updated raw reduction that converts from 2-dimensional to 1-dimensional spectra \citep{Howard2010}. This work improves the quality of the S-values compared to CLS1 by using a restricted extraction width of eight pixels to reduce sky-emission and scattered light contamination and by making additional quality controls.  A sample of Ca H-line profiles for properly reduced spectra, in good seeing are shown in Figure \ref{fig:hline_profile}.

HIRES spectra are collected with a variety of decker apertures. The primary science deckers for CPS are B5 (0.87'' x 5'') and C2 (0.87'' x 14''), which provide a resolution of 60,000. Two other deckers, the B1 (0.5'' x 5'') and B3 (0.5'' x 14''), are used for templates and result in a resolution of 80,000. 
The C2 decker began use in 2009 June when the typical visual magnitude of RV targets changed from V$\sim$8, for nearby star surveys, to V$\sim$12, for follow-up of \emph{Kepler} planet-host stars. Sky contamination became a limiting factor in RV precision, requiring observations with the C2 decker \citep{Marcy2014}, at the occasional expense of useful S-value measurements. The B5 decker was used for stars brighter V $\sim$ 10, and the C2 was used for fainter stars and for observations taken during twilight, when CLS1 stars are often observed.

\begin{figure} 
\includegraphics[width = 1.0\columnwidth]{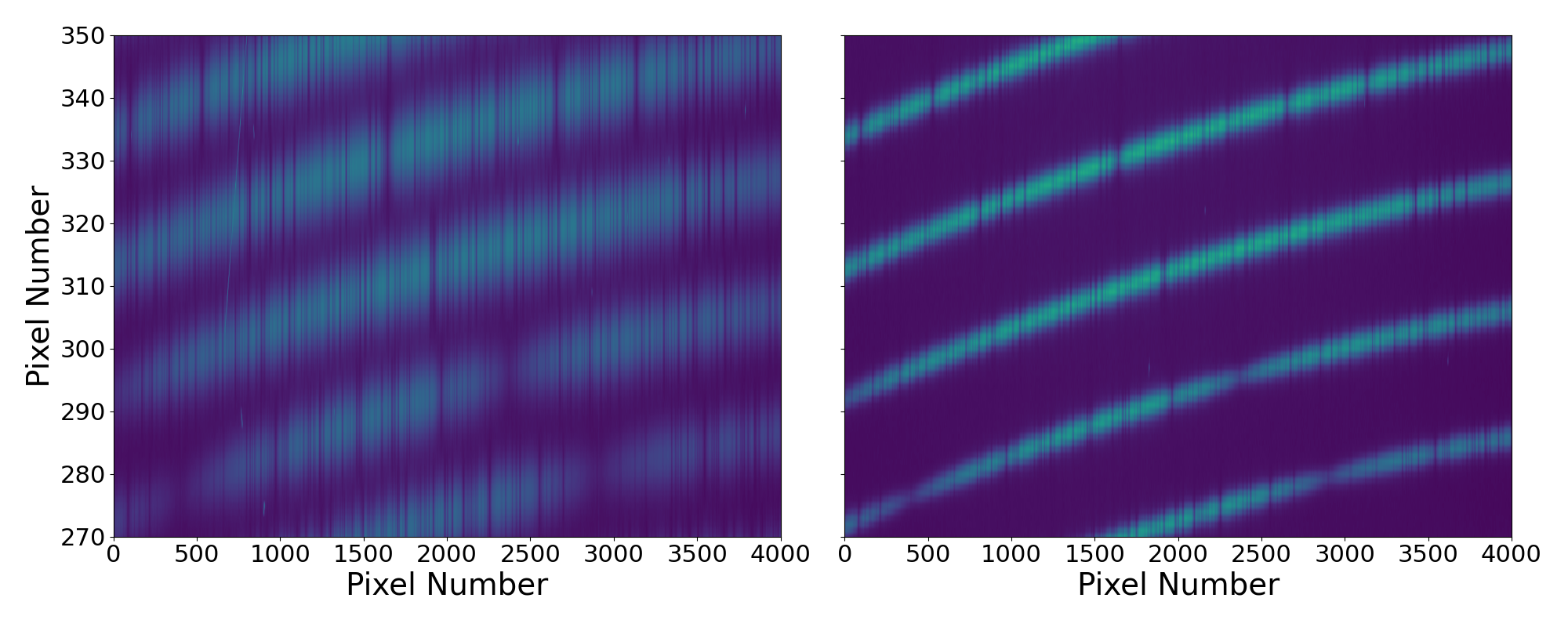}
\caption{ 
Two raw images of HD 141399 showing the \cahk region. Left: Seeing measured to be 2.5", Right: Seeing measured to be 0.9".   
The extraction width in the cross-dispersion(vertical) direction is limited to eight pixels.}
\label{fig:raw_data} 
\end{figure}

The height of the C2 and B3 deckers allows for simultaneous observations of sky pixels and causes order overlap on the middle CCD, and increasing overlap blueward. Echelle spectrographs with cross dispersing gratings have blue orders closer together and red orders with larger separation, and the opposite is true for cross dispersing prisms. The raw reduction has been tailored to account for this in the middle detector, resulting in equal RV precision for the B5 and C2. For S-values, the additional overlap near the \cahk lines is more problematic and causes degraded quality for observations taken in poor seeing conditions. By measuring the stellar profile in the spatial direction, we calculate the average seeing for each observation. Using chromospherically inactive stars, we have identified the upper limit of 1.6'' to be the critical seeing value required to avoid order-to-order contamination when observing with the C2 or B3 deckers \citep{Baum2022}.  We exclude S-values with seeing measurements beyond this value from our sample, removing 1549 S-values. The B5 and B1 deckers with their shorter height do not have order overlap and do not have this restriction. 
Figure \ref{fig:raw_data} shows a 2D echellegrams for the star HD 141399 taken in seeing conditions of 2.5'' and 0.9'', showing the order overlap that occurs during poor seeing conditions. In addition to the quality control described above, we visually examined exceptionally low S-values and excluded 98 observations with poor extractions.

\subsection{S-value Error}
We previously adopted the S-value error of 0.002, or 1\% per individual observation, by assessing the S-value distribution of HD 10700 (Tau Ceti), a star with well-established low level of activity \citep{Isaacson2010}.  Our extended time baseline yields a dispersion of the HD 10700 S-values of 0.82\% (0.00139/0.1675). We identified other stars with very low S-value variation, including HD 55575, our least active star that has a dispersion of 0.0007/0.156 = 0.45\%. We adopt an S-value error of 0.001 for all observations, a value between the dispersion of HD 10700 and HD 55575. The HARPS-AMBRE survey found a dispersion of 0.83\% for HD 10700, showing that this precision level is achievable \citep{GomesdaSilva2021}. We discuss the least active stars in Section \ref{sec:leastactive}.

\subsection{Sampling}
The CLS1 survey required 10 RV observations over 8 years, since 2005, on HIRES (Figure 2 in CLS1) to be included in its analysis. They supplemented their dataset with pre-2005 HIRES RVs and Lick Observatory RVs \citep{Fischer2014}. We do not include the Lick Observatory S-values nor the pre-2005 HIRES S-values. We considered adding S-values from the Automated Planet Finder (APF), but there is no additional baseline since the first observations were taken in 2014. Since CLS1 we collected four additional years of S-values, improving the baseline for many stars. 

We are primarily focused on finding stellar activity cycles with periods between 2 and 25 years, so we require 45 observations since 2005. Stars with fewer, often sporadically timed, observations are insufficient for robustly detecting activity cycles. We include stars with as few as five measurements for the average activity analysis.  
Out of 710 stars in our sample, we search 285 for cycles, and the 425 additional stars are included in the summary activity analysis.

\subsection{Data Validation and Rejection} 

To ensure the highest quality dataset, we start by adding S-values for non-iodine observations that were omitted from CLS1 because they do not contribute to the RV time series. The spectral segments used to calculate the S-values are shifted and scaled to a high signal-to-noise (SNR) template of that same star \citep{Isaacson2010}. For eleven stars, no such template exists, so we use a spectrum of Vesta, a reflective spectrum of the Sun, that is shifted to observatory wavelength solution. Vesta spectra have the benefit of having the same format, and blaze function as all other HIRES spectra. Those stars are HD 114762, HD 120136, HIP 60633, HD 152391,  HD 10853, HD 6101, HD 112914, HD 167042, HD 73344, HD 177153, and HD 8375. S-values for these stars have the same quality and uncertainties as the others.

We make the following requirements at the level of individual observations for quality control.

\begin{itemize}
   \item S-values less than 0.10 are rejected as non-astrophysical. Fifty-six S-values are removed, and 98 are identified by-eye as having poor extractions.
   \item The SNR must be greater than 7 at continuum near 4000 \AA, removing 276 S-values. One star, GL 406, has no observations that meet this threshold and is omitted throughout our analysis. See \citep{Bowens-Rubin2023} for a detailed analysis of this star.
   \item The seeing must be less than 1.6" for C2 and B3 observations, excluding 1549 S-values from our sample. 
   \item  Stars with \teff < 4000K were visually inspected and stars with \cahk activity that extends beyond the 1.09 full width half-max \AA~ window that is used to calculate S-values were removed. Eight flare stars have flux emission in the H\&K region that is not well modeled using S-values are excluded. These stars also have a spectral helium line in emission that resides very near the H line, causing further ambiguity in the S-value measurement. The stars excluded are GL 83.1, GL 876, GL 905, HIP 112460, HIP 37766, HIP 5643, HIP 92403, and HD 75732B. 
   \item We retain 52372 S-values from 2005 through 2023 October for 710 stars.
\end{itemize}

\begin{figure*} 
\begin{center}
\includegraphics[width = 2.0\columnwidth]{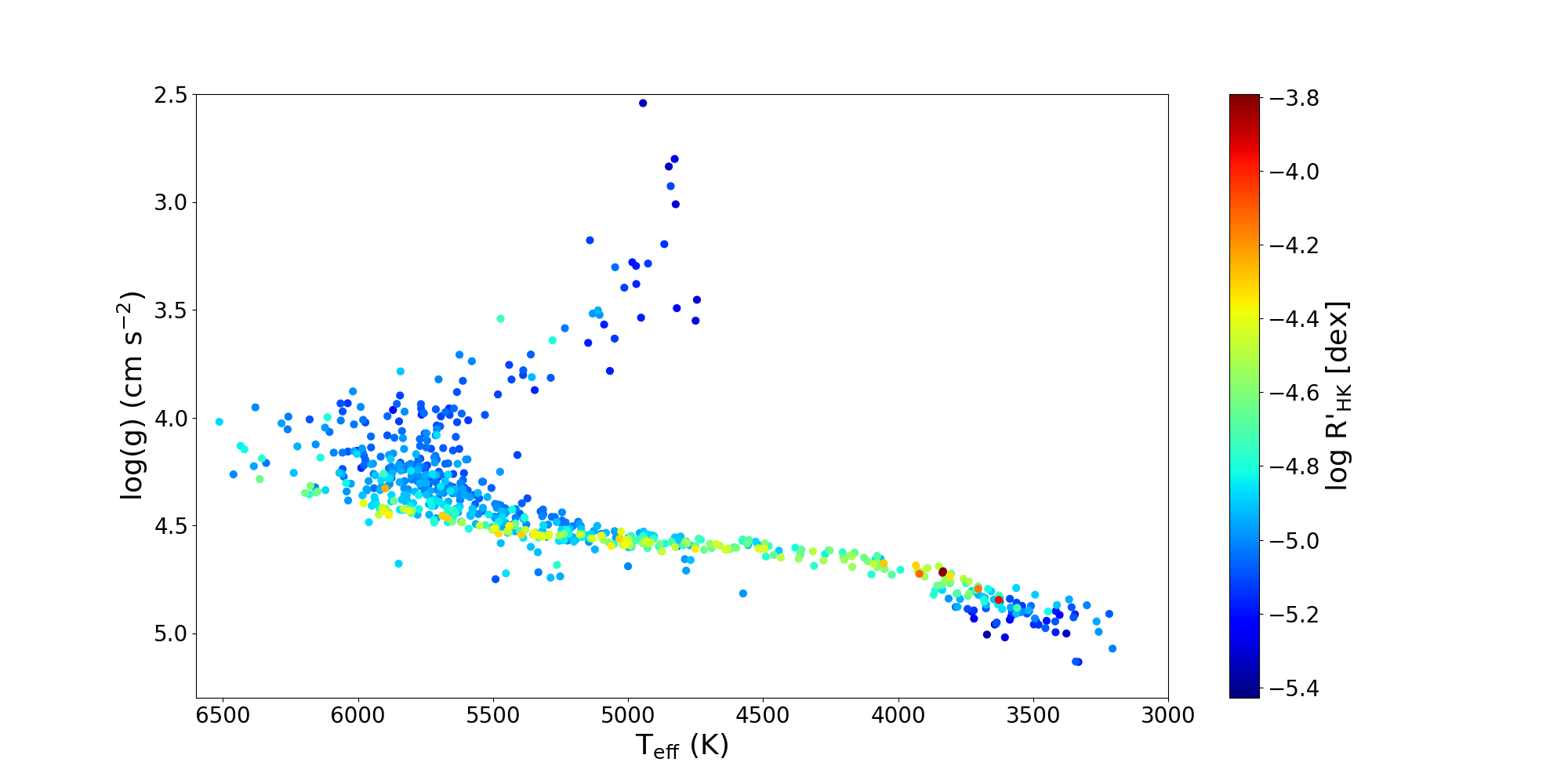}
\end{center}
\caption{Stellar surface gravity as a function of effective temperature is plotted, with the median \rphk value for each of the 710 stars in the CLS Activity Sample on the color scale. Subgiants are visible at lower \logg values, while the few very low-metallicity subdwarfs in our sample fall below the main sequence. Most stars in our sample are slowly rotating FGKM stars on the main sequence.}
\label{fig:teff_logg_rphk} 
\end{figure*} %

\section{The 710 Star Activity Sample} \label{sec:activity_sample}

\subsection{Overview}
To assess the average activity of our sample we begin with the 710-star CLS1 sample that consists of slowly rotating FGK and M type stars that are amenable to RV measurements in search of exoplanets. Figure \ref{fig:teff_logg_rphk} presents our sample in the \logg vs \teff plane showing the average activity, \rphk, as the color scale. The sample was assembled to offer consistent sensitivity to long period giant planets out to tens of astronomical units. The minimum baseline chosen of 8 years, with 10 observations from HIRES and 20 total RVs, complements our search for stellar activity cycles, which tend to range from 2-25 years \citep{Baliunas1995,Baum2022}.  With many observations spanning the timescale of typical activity cycles, we present accurate measurements of the average activity for each star.

The CLS sample was originally selected to exclude stars that host known transiting planets and stars with known high metallicity. Samples of stars that focused on sub-giants, young stars, and those that had long baseline observations due to the presence of hot Jupiters were also excluded. There are 178 known exoplanets or brown dwarfs around the stars in our sample.

\subsection{Stellar Property Corrections}
We amend the stellar properties catalog from \cite{Rosenthal2021} by filling parameter values for five stars lacking \teff. For HD 134439 and HD 134440, two chemically peculiar twin stars in a binary system with an long period orbit, we use the \teff, \logg, \feh and \mstar from \cite{Chen2014}. For HD 201092, and HIP 106924, we add \teff from our SpecMatch-Synthetic analysis of HIRES spectra \citep{Petigura2017}. We obtain the extremely low metallicity value of -2.5 for HIP 106924 from \cite{Joyce2018}. For GL 528 B, we apply the SpecMatch-Empirical code to a HIRES spectrum \citep{Yee2017}. These are important additions since we are interested in the dependence of activity on \teff.

\subsection{Derived Properties}

We calculate \rphk \citep{Noyes1984} and stellar age \citep{Mamajek2008} to examine activity correlations with fundamental and derived stellar properties and to check for correlations with activity cycle properties. \cite{Cincunegui2007}, \cite{suarezmascareno2015}, \cite{AstudilloDefru2017}, \cite{Mittag2013} and \cite{Marvin2023} each extend the \rphk calibration to cool stars (B-V = 1.6) and \teff = 2700 K. We use the \cite{Noyes1984} method to enable cross referencing with other works, rather than \cite{Marvin2023}, which overestimated the color correction factor resulting in overestimation of \rphk. 

Since B-V colors are used to calculate \rphk, we derive B-V using \cite{Ramirez2005} which uses both \teff and \feh and is valid from \teff of 7000 - 3870 K. For stars cooler than 3870 K, we use the CLS1 B-V values.   As a result, we  use the \cite{Noyes1984} method for all of our analysis of \rphk, and derived stellar parameters. The use of consistent \teff values when converting to \rphk will be critical to our cycle period to activity analysis. With this choice, we urge caution when using \rphk values for stars cooler than 3870 K due to the uncertainty in B-V.

\subsection{The 710 Stellar Activity Analysis} 
\label{activity_sample}
Our stellar sample is presented in temperature vs surface gravity space in Figure \ref{fig:teff_logg_rphk} with a color scale indicating the stellar activity, \rphk. The prominently positioned sub-giant stars that rise above the main sequence have the lowest stellar surface gravity values. Sub-giant stars, with their larger stellar radii typically have less stellar activity than main sequence stars of the same \teff. As they evolve and expand, their rotation rate slows to conserve angular momentum, and the decrease in density produces a more subdued stellar dynamo. 

The zero-age main sequence is visible as active stars with high \rphk. Sub-dwarfs, with  extremely low metallicity and old ages, fall below the zero-age main-sequence. The coolest stars in our sample, below 4000 K, have a variety of activity values including the eight eruptive variables which we exclude. This may be due to the \rphk metric being calibrated by \cite{Noyes1984} on sun-like stars. \rphk values are also sensitive to the choice of conversion from \teff to B-V.  Using the average of the stellar activity time series makes for a robust measurement of the average activity of our sample. Each of the stars in our sample has at least 5 observations.

\begin{figure} 
\includegraphics[width = 1.0\columnwidth]{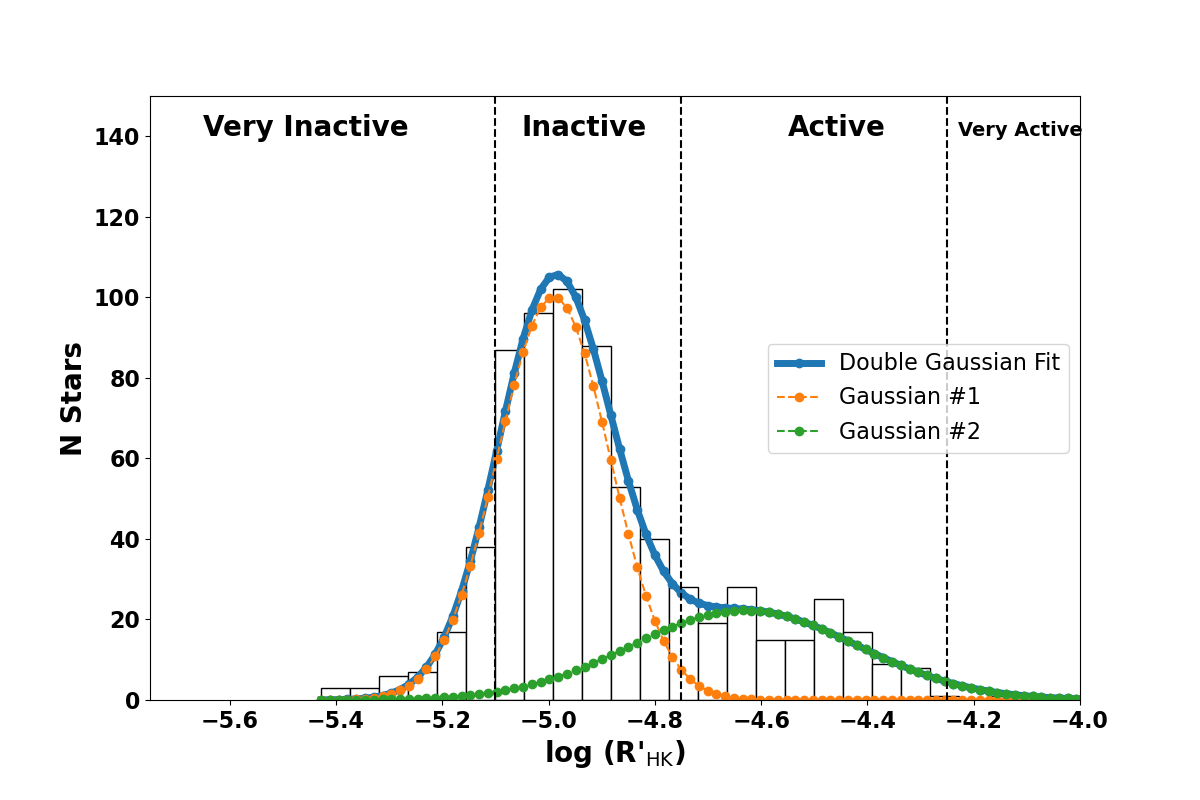}
\caption{ 
The distribution of our \rphk data is modeled with a double Gaussian model with peaks at $-5.001$ and  $-4.886$. We find less structure than \cite{GomesdaSilva2021}, perhaps due to the smaller number of stars in our survey compared to theirs. Activity qualifiers come from \cite{Wright2004}. All 710 stars are included here.}
\label{fig:hist_2gaussian} 
\end{figure} 

\begin{figure} 
\includegraphics[width = 1.0\columnwidth]{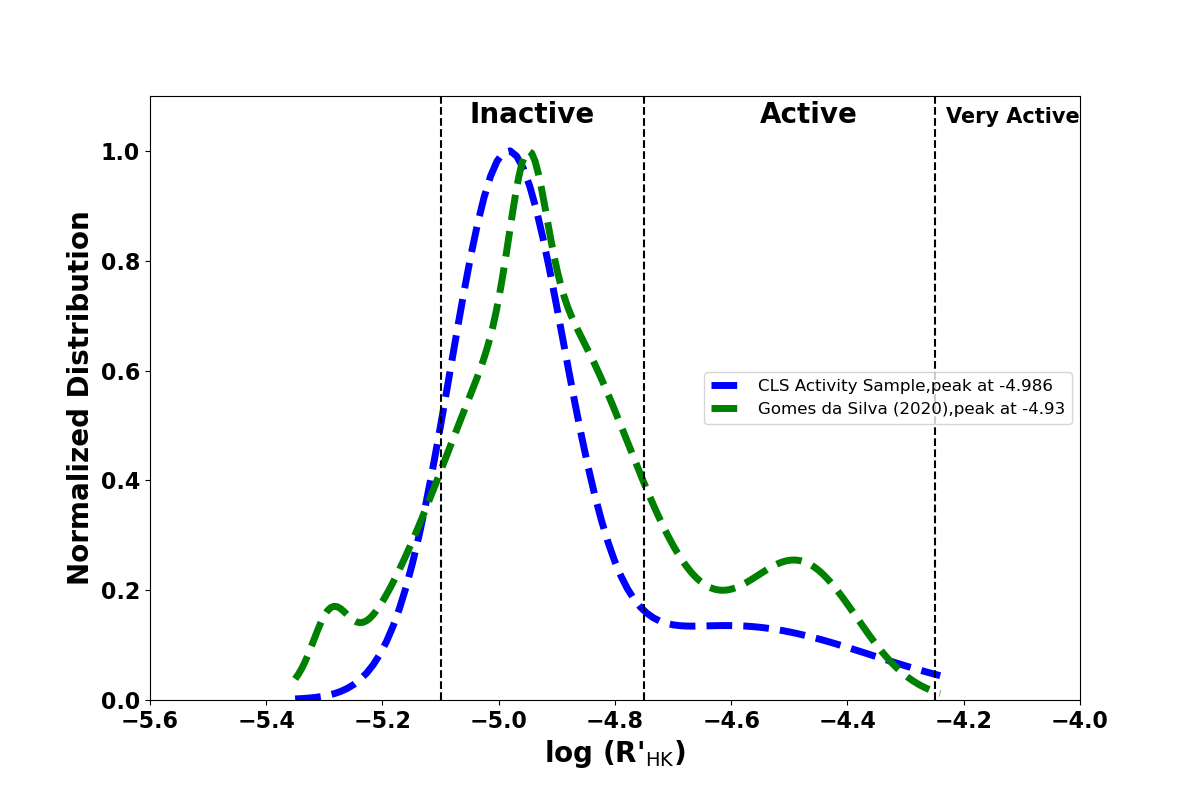}
\caption{ 
The distribution of our \rphk data is modeled with a double Gaussian model and compared with \cite{GomesdaSilva2021} which had a \teff lower limit of 4500 K. Our sample extends to \teff of 3000 K so we omit stars cooler than 4500 K from this plot to make the comparison more direct. We see the two main peaks show up in both our sample and \cite{GomesdaSilva2021}. Our main peak is slightly offset in the direction of less activity. \cite{GomesdaSilva2021} models four Gaussians to our 2 Gaussians, potentially causing this offset. The amplitude of the offset is 0.10, or twice the calibration offset for S-values from different instruments \citep{Mittag2013}. The astrophysical explanation is that CLS gave lower priority to more active stars.} 
\label{fig:hist_compare_samples} 
\end{figure}

We compared our sample to the primary southern-sky planet search survey, the AMBRE-HARPS survey \citep{GomesdaSilva2021}, by modeling our distribution of \rphk values as a sum of Gaussian contributions. Their catalog contains 1674 planet search stars, and they also focus on slowly rotating F, G and K-type stars. Figure \ref{fig:hist_2gaussian} shows our two-Gaussian fit to our 710-star sample and Figure \ref{fig:hist_compare_samples} shows our sample compared to the AMBRE-HARPS sample. We normalize the y-axis and remove stars cooler than 4500 K in this plot to make a direct comparison as possible. Their sample shows more structure than ours, and peaks at a slightly more active value. Their small peak near -5.3, attributed to giant stars is not present in ours, due to our lack of giants. The additional structure for active stars may be due to their larger sample or is perhaps due the CPS observing strategy of excluding active stars at early points in the survey. One possible systematic difference is the slit-fed vs fiber-fed spectrographs, HIRES and HARPS, respectively. However, we think this is sufficiently addressed in Section \ref{sec:data_quality}. The Gaussian properties representing our sample are available in Table \ref{tab:table_fits}.

\begin{figure} [ht]
\includegraphics[width = 0.9\columnwidth,trim={0cm 1cm 0cm 1cm},clip]{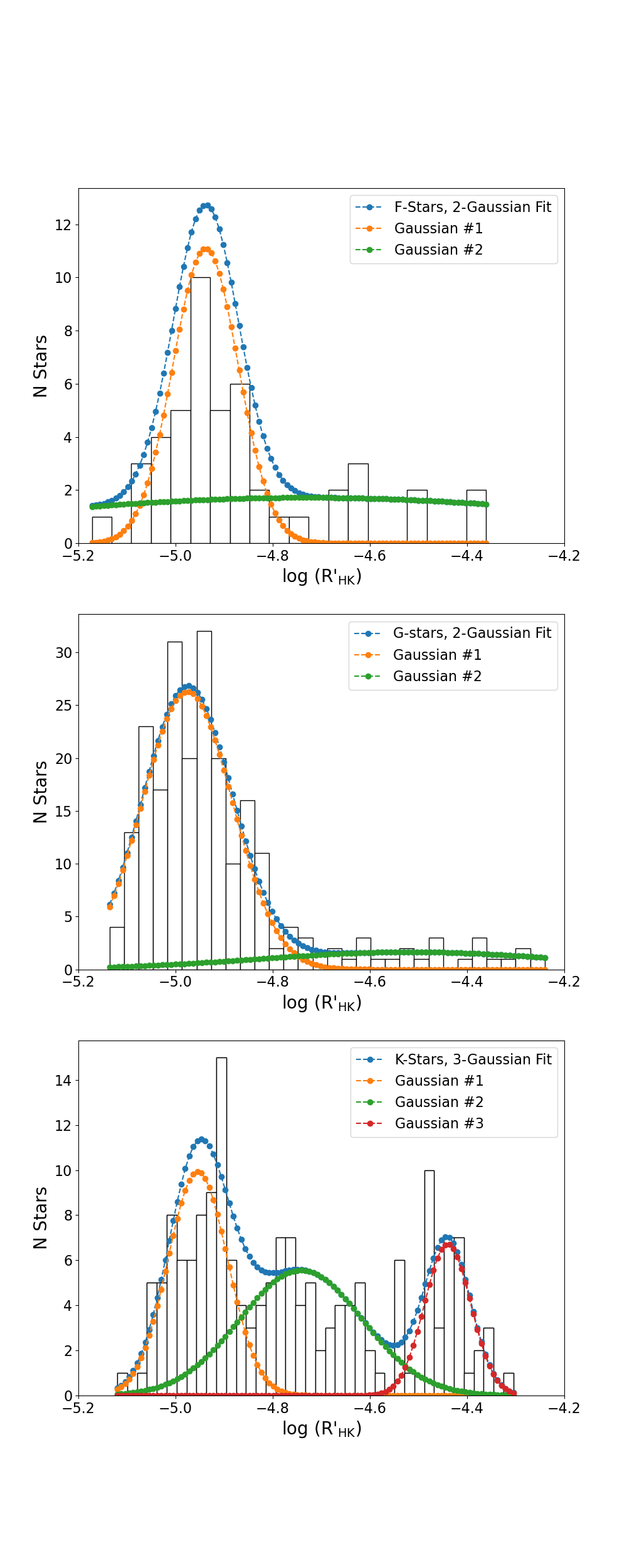}
\caption{From top to bottom, the distributions of the \rphk for the CLS sample are shown for F-stars, G-stars and K-stars, with \logg > 4.2. We model each distribution with a series of Gaussians and compare them to \cite{GomesdaSilva2021}. Structurally the sample agrees with the AMBRE-HARPS sample of planet search stars from the southern hemisphere.} 
\label{fig:hist_FGKstars_gaussians} 
\end{figure}

We constructed histograms for each of the FGK stars to compare the two samples as a function of spectral type (Figure \ref{fig:hist_FGKstars_gaussians}). The smaller number of stars in our sample means that features in each distribution are not well defined, which leads to several degenerate model fits for each spectral type. Structurally the comparisons for each type of star are quite similar. The F-stars have a broad distribution, represented by two Gaussians at \rphk of -5.018 and -4.934.
The G-star distribution is dominated by the primary peak near -5.0 and a secondary peak of more active stars, at -4.50, which is broad and low-amplitude. Our distribution of \rphk for K-stars is distinctly triple peaked, similar to \cite{GomesdaSilva2021} but with a smaller peak at higher activity. Overall, we find the by-type comparison to be consistent with that found for planet search stars in the southern hemisphere.

\subsection{Time-Averaged Activity}
\label{sec:activity_variability}

\begin{figure*}  
\includegraphics[width = 6.9in]
{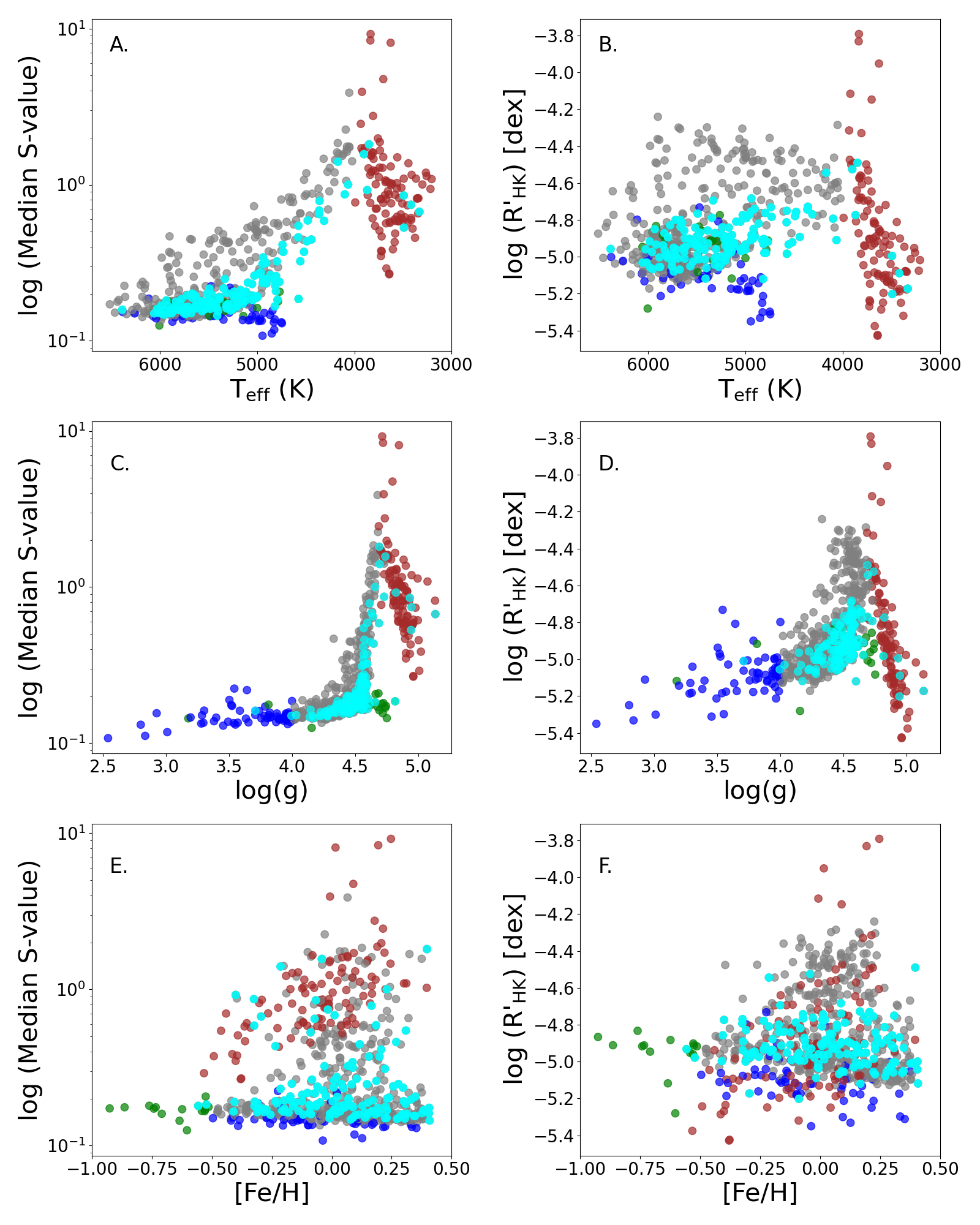}
\caption{ 
The values of \teff (panels A. B.), stellar surface gravity (panels C. and D.) and \feh (panels E. and F.) are plotted versus S-value and \rphk for 710 stars. Maroon data points represent cools stars, with \teff < 4000 K. Blue data points are stars with \logg < 4.0, sub giants, and green data points represent stars with metallicity less than -0.5. Cyan symbols identify stars with cycles. The bottom two panels exclude 11 stars with \feh < -1. Grey data points represent stars not in the extremes of \teff, \logg, and \feh and without cycles.}
\label{fig:teff_logg_fe_sval_rphk_v2} 
\end{figure*}

We plot the S-value and \rphk of the 710-star sample against fundamental stellar properties, highlighting, with different colors, the extreme ends of each stellar property distribution (Figure~\ref{fig:teff_logg_fe_sval_rphk_v2}). Maroon data points represent stars with \teff < 4000 K. Blue data points represent evolved stars, with \logg < 4.0. Low-metallicity stars with \feh < -0.5 are represented in green. Cyan x's identify those stars with cycles identified in Section \ref{sec:cycles_sample}.

In panel A. \teff vs S-value are plotted and the familiar flat floor of activity from 6500 K to 5000 K is visible, followed by a steady increase from 5000-4000 K \citep{Isaacson2010,Mittag2013}. The slope of the activity floor inverts down to our lowest \teff stars. There is a lower density of stars from 6000-5000 K that is elevated above the primary distribution of very inactive stars. The low-metallicity stars and low \logg stars fall near the S-value floor, as expected for older stars and sub-giants. Panel B, with \rphk as a function of \teff shows that most stars in our sample are between 5000 K and 600K.  The more active stars that lie above -4.8 may have cycles but too variable to be strictly periodic.

Panel C. plots median S-value as a function of stellar surface gravity showing the low gravity stars that have started to evolve off of the main sequence are at the floor of the S-values distribution. The highlighted very-low metallicity stars mostly fall in a unique parameters space at a \logg of 4.6-4.7. Although our metallicity distribution is sparse at \feh < -0.5 dex, the difference in S-value at that specific \logg is distinct. The elevated population of stars at \teff between 5000-6000 K in panel A is now compressed in panel C. at a \logg value of 4.5 to 4.7.  This could be related to the age of these stars with younger stars being more active and having higher \logg. Panel D shows the most active stars to have \logg near 4.5, indicating that they are near the zero-age main sequence.

Panel E. plots the median S-value for 710 stars as a function of metallicity. At the bottom of the distribution, the subgiant population spans a wide range of \feh.  Our sample is slightly overrepresented at \feh greater than solar (45/55\%), but is sufficiently populated from $\pm$0.4 such that we can draw conclusions about the presence of stellar activity cycles as a function of metallicity in Section \ref{sec:cycle_search}. A Pearson correlation coefficient between average S-value and metallicity and also S-value RMS with metallicity was calculated and found to be less than 0.1 in each case. This suggests there is weak correlation between spectroscopic activity metric S-value and metallicity, in contrast to photometric correlations to \feh such as noted in Kepler flare stars \citep{See2023}. Our findings are consistent with \cite{Lovis2011} in which the B-V to temperature conversion accounts for metallicity.

\subsection{Activity Variability}
\label{sec:activity_variability2}

\begin{figure*}      
\includegraphics[width = 2.0\columnwidth]{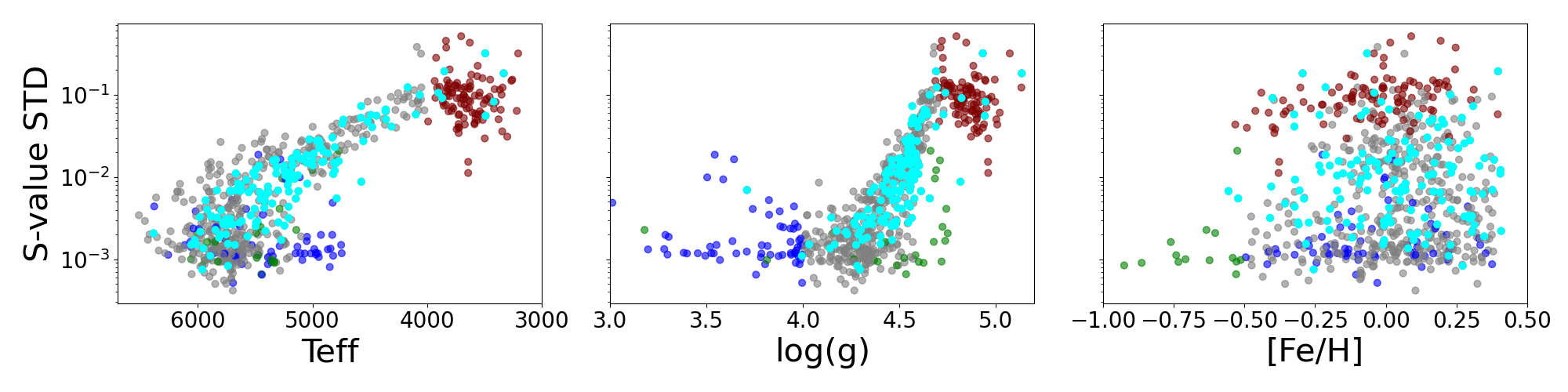}
\caption{S-value standard deviation is plotted as a function of \teff, \logg and \feh for all 710 stars in our full sample. Stars with low \logg are in blue, and stars with \teff < 4000 K are maroon. The lowest-metallicity stars shown in green. Eleven very low-metallicity stars are off of the plot. Cyan marks stars with detected cycles. The dichotomy of stars with low \logg values is intriguing. We do not sample all of the parameter space equally, but the marked cycles still identify the area most likely to find periodically cycling stars. The grey data points represent stars with \teff > 4000 K, logg > 4.0, and \feh > -0.5 and include star stars above and below our threshold of 45 observations.}
\label{fig:teff_logg_fe_sval_variation} 
\end{figure*}

The long time baseline over which these observations were collected and the multiple observations for each star leads to robust measures of average activity of our sample. We examine the variability of S-value as a function of \teff, \logg, and \feh in Figure \ref{fig:teff_logg_fe_sval_variation}. Stars below 4000 K are the most variable, with significant variation due to the stellar rotation period variations. These stars are more heavily spotted, confusing our search for sinusoidal stellar activity cycles with periods on timescales of years. We adopt the values of \teff, \logg, \feh, \mstar, and \rstar from \cite{Rosenthal2021}, except where noted in Section \ref{sec:data_quality}. Table \ref{tab:table_average_activity} contains the minimum, median, and maximum S-values, S-value root-mean-square (RMS) value, standard deviation, \rphk,  \rphk RMS, number of observations, and activity derived stellar ages. The S-value times series are provided in Table \ref{tab:table_time_series}.

\section{The 285 Star Sample to Search for Activity Cycles}
\label{sec:cycles_sample}

We identify robust stellar activity cycles for use in analysis of stellar cycles as a function of stellar properties, including age. By requiring 45-observations, we define our Stellar Activity Cycle sample of 285 stars. This choice helps to avoid a spurious detection of an activity cycle due to poor sampling.

\subsection{Searching the 285 Star Sample for Activity Cycles}
\label{sec:cycle_search}

The patterns of stellar activity can have complex structure on many different timescales, so we choose a simple sinusoidal model, with no eccentricity, to search for signals with periods between 2 and 25 years. Cycles less than two years are difficult to identify due to seasonal sampling, and ambiguity with rotation periods \citep{Baliunas1995,BoroSaikia2018}. We expect only the youngest (Age less than 1 Gyr) to have such short stellar activity cycles and the CPS RV planet surveys tend to exclude young stars in blind surveys. Notably, HD 115043 has a stellar activity cycle of 1.7 years \cite{BoroSaikia2018}, but is not in our sample. HD 22049, Epsilon Eridani, has a multiple previously published cycle of 2.2 and 12 years \cite{Metcalfe2013} or perhaps a 3-year, 11-year and 34-year cycle \cite{Fuhrmeister2023} when the calcium infrared triplet and X-ray measurements are analyzed. We have the sensitivity to detect these cycles with our data, but none of these periods pass our threshold. While the zero-eccentricity sinusoid model is sufficient for uniformly identifying activity cycles, a more complex model should be chosen for when modeling young star cycles and those with complex signals such as HD 18803, HD 219134, HD 201092, and HD 140538A.

We utilize the Lomb-Scargle periodogram routine in Astropy\footnote{ \href{https://docs.astropy.org/en/stable/timeseries/lombscargle.html} {https://docs.astropy.org/en/stable/timeseries/lombscargle.html }Astropy's  LombScargle Periodogram, see section on normalization='model'} 
and the 'model' normalization option to identify peaks in the periodogram and fit a periodic function to the tallest peak. The 'model normalized' periodogram normalizes the periodogram around the residuals to the periodic model, rather than the constant model that is the default method of normalization. This normalization also accounts for the offset from zero typically expected from the generalized Lomb Scargle periodogram. We limit the periodogram from 100 to 10000 days (27.4 years) and explored alternate limits such as 200-2000 days with no effect on recovered activity cycles. 
The strength of this method is the simplicity of the model that is easy to parameterize and search. The weaknesses are that stellar activity cycles do not always stay in phase over many cycles, some cycles are better modeled by adding eccentricity, and stars with multiple cycles and different periods are difficult to identify.

We combine two quantitative metrics to identify stellar cycles. First, we calculate the difference in the standard deviation of the initial S-values to the standard deviation of the residuals, divided by the initial S-value standard deviation and the median S-value for that star (Equation \ref{equation1}). Our threshold by this metric for detection is 1.20. We attempted to use Chi-squared as best-fit metric but found the scatter in S-value over timescales of weeks and months makes this metric less useful. We determined the threshold value by ranking our stars with this metric and finding where the cycles become unreliable by eye. 

\begin{equation}
\label{equation1}
\mathrm{Threshold} = \frac{(\mathrm{STD}_f - \mathrm{STD}_i)}{ \mathrm{STD}_i * \mathrm{Median (S-value)} } 
\end{equation}

As a secondary metric, we identify cycles through the Lomb-Scargle periodogram as those having a peak greater than 0.5. The maximum peak of any star is 14.0 (HD 192310) and the median peak value for all detected cycles is peak is 2.01. The weakest signal in our data if GL 699, with a peak value of 0.501. We also remove stars with candidate cycles if the second peak in the periodogram is more than 75\% of the primary peak. Such peaks indicate that the identified period is not sufficiently unique for our purposes. Using the ratio of the first and second tallest peaks eliminates many stars that have very plausible cycles but have ambiguous periods. Many stars with cycles much longer than our observing baseline are removed in this quality cut. With the periodogram qualification, we recover 27 cycles that do not pass our threshold from Equation \ref{equation1}, including all of the detected cycles in stars below \teff of 4400 K. By including our reliability metrics for all 285 stars, future studies can choose different thresholds to fit their analysis needs. 

For three stars, we move a linear trend and afterward detect an activity cycle. These trends are indicative of a second cycle in the same, as has been studied most recently by \cite{Mittag2023}. Those stars are HD 219134, HD 158633, and HD 82943. When a trend is removed for HD 23356 and HD 201092, they show candidate cycles but are do not pass the quantitative thresholds. Generally, our search method is insensitive to cycles longer than our baseline of 20 years and fails to robustly identify any previously unknown secondary cycles.  

These combined metrics detect stars with incredibly small overall variations such as HD 126614 with a peak to peak S-value amplitude fit value of 0.0041. For comparison, HD 10700, considered an activity standard has  an S-value standard deviation of 0.0012 and relative dispersion of 0.8\%.  The least active stars with and without cycles are discussed in Section \ref{sec:leastactive}. From 285 stars searched, we present 138 stellar activity cycles and next discuss our recovery of previously known cycling stars.

\subsection{Cycle Comparison to Previous Studies}

\subsubsection{Mt. Wilson Cycles}

Most published activity cycles that produced with S-values collected by the Mt. Wilson Observatory H\&K Project.  \cite{Baliunas1995} found that 52/111 main-sequence stars have cycles. The regularity of stellar cycles is dependent on age ``young stars rarely display a smooth, cyclic variation...intermediate age star have occasional smooth cycles...stars as old as the sun have smooth cycles'' \cite{Baliunas1995}. Although our sample is vastly different, we find 138/284 stars have cycles. Our sinusoidal search is not sensitive to non-periodic cycles of young stars such as HD 22049. Our search is most sensitive to the regular cycles of older than 1 Gyr. The largest sample of Mt. Wilson stars compiled has 335 stars. From our 710 sample, 173 stars are overlapping and we independently identify 44 cycles from those 173 stars. Since our sample contains a broader range of stellar types, we are exploring parameter space beyond \cite{Baliunas1995} allowing for the examination of correlations between cycle periods and stellar properties.

Summary analysis of the Mt. Wilson cycles has claimed that the sun is near the upper mass limit for cycling stars \citep{Schoeder2013}, but the identification of many cycles down to 4400 K presented here, and a handful of M-dwarfs found here and by \cite{Irving2023} show that stars across the main sequence can be in cycling states.

\subsubsection{Studies of Keck data}

We cross-checked the 13 cycles from 59 stars examined in \cite{Baum2022}, adding two years of HIRES data, and find 10 of 13 cycles. Of those not detected, HD 166620 is in a well-studied minimum phase of its cycle \citep{Luhn2022}, while HD 101501 and HD 152391 have fewer than 45 observations. For the cycles that we expect to detect, our algorithm identified them.

The CLS activity time series has been analyzed in relation to precise RVs.  In CLS1, Table 7 mentions 43 false positive planet signals that are attributed to stellar activity cycles, and 13 false positives due to rotation periods. We confirm 41 of 43 as stellar activity cycles. \cite{Luhn2020} lists stars with possible stellar activity cycles, but the focus of that work was on average activity and its impacts on RV precision. For their stars that overlap with our sample, we quantify the intensity and period of those signals. 

\subsubsection{Johnson et. al. (2016)}
\cite{Johnson2016} use data collected 295 spectra from the 2.7 m Harlan J. Smith Telescope at McDonald Observatory from 1998 to 2015 to monitor the radial velocity of the HD 219134 system and also collected S-value measurements of stellar activity. They detected a stellar activity cycle of 11.7 years with no linear trend. Our analysis finds a cycle period of 13.4 years after removing a robust linear trend over the observation baseline of 20 years. Understanding the discrepancy in the detection of the linear trend may be explained by differing measurement uncertainties and is worthy of further exploration.

\subsubsection{Toledo-Padron et. al.(2019)}
\label{sec:gl699detect}
\cite{ToledoPardon2019} studied the stellar activity of GL 699, Barnard's star and revealed an activity period of 8.8 years in the \cahk S-values. Our measured cycle of 8.5 years is consistent with their value. \cite{ToledoPardon2019} also use photometry to determine the activity cycle finding a  10.5 yr cycle. Their conclusion that GL 699 is a very inactive star is consistent with our finding of the periodogram amplitude being just above the threshold of detection.

\subsubsection{Mittag (2023)}
\cite{Mittag2023} list 34 chromospherically detected activity cycles around FGK stars. For the 15 stars that overlap with our sample, and have more than 45 observations, we recover 14 of the cycles. HD 201092 has two cycles, but neither passes our thresholds. The 11 year period is not detected due to our choice to fit only circular periodicity. Future studies that include the Mt. Wilson and Keck/HIRES datasets can potentially confirm these complex cycles. Future analyses may require different levels of confidence in the cycle detection or will be conducted with a different model.

\subsubsection{Studies of Fully Convective Stars}
\label{sec:fullyconvective_study}

Photometric data from the ASAS-SN project identified 13 of 15 fully convective M dwarfs showing stellar cycles \citep{Irving2023}. For overlapping stars in our sample with 45 observations HIP 80824 (GJ628) and HIP 109388 (GJ849) are not detected. GJ 317 has two cycles, we detect neither. HIP 103039 (LP 816-60), HIP 57548 (GJ 447), GJ 285, GJ 54.1, GJ 234, and GL 406 all have fewer than 45 Keck/HIRES observations, falling below inclusion threshold. The cycles with periods of a few years that \cite{Irving2023} detect with photometry must not have periodic signals amenable to detection with the \cahk emission lines.  The lack of chromospheric confirmation of the photometric cycles may be due to complex structures to which our simple sinusoid model is not sensitive. Indeed other studies have identified chromospheric cycles in M-dwarfs \citep{suarezmascareno2016,Wargelin2017}.

From this point on we turn our focus to the 285 stars that have at least 45 Keck/HIRES S-values since 2005, eventually narrowing down parameter space to show that nearly every star in that range has a periodic activity cycle.

\begin{figure*} [ht] 
\includegraphics[width = 2.1\columnwidth]{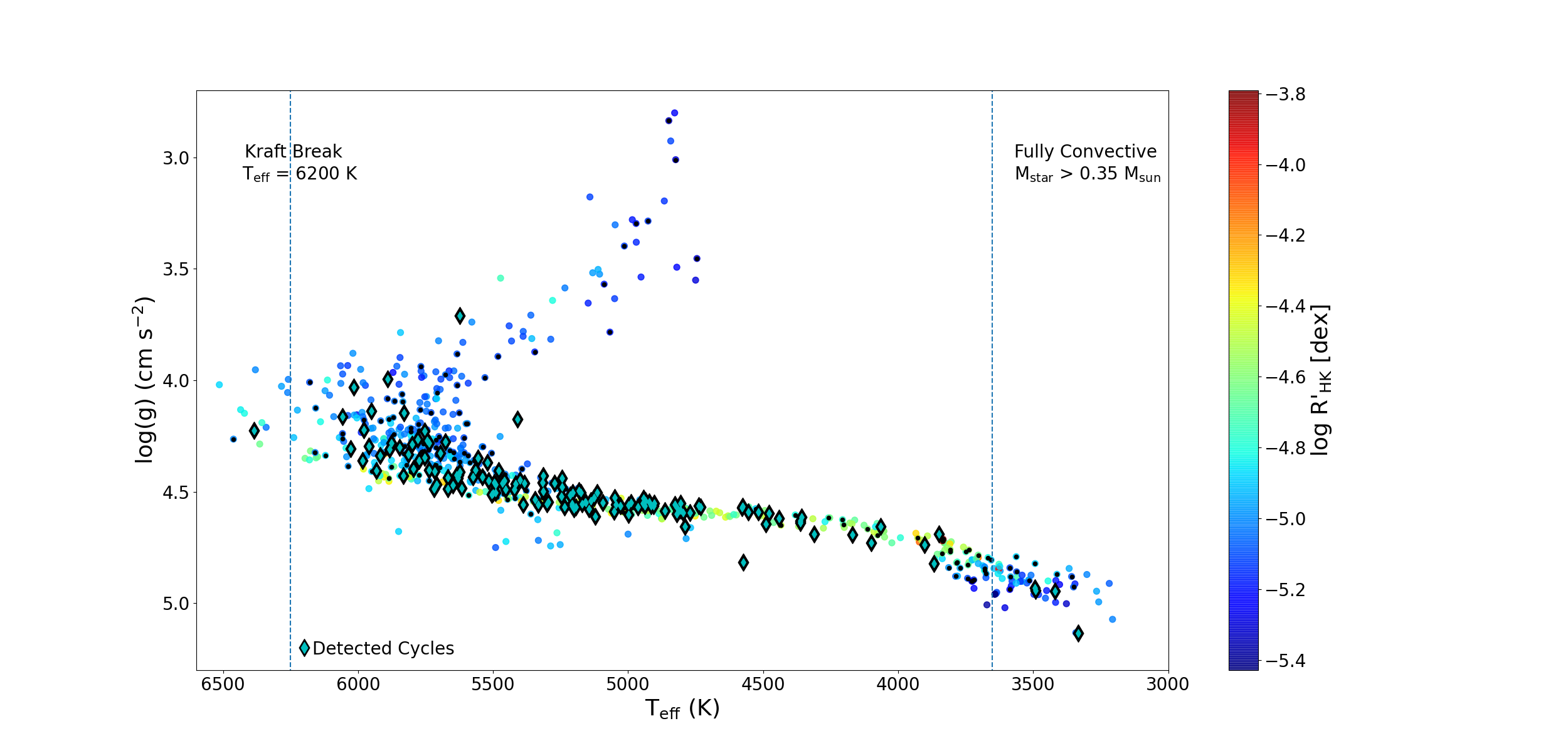}
\caption{ 
Effective temperature, stellar surface gravity and median \rphk value for 710 stars in our sample. Diamonds show detections of 138 stellar activity cycles. Circles identify the 285 stars with more than 45 observations that were searched for cycles, and the color scale represents \rphk.}
\label{fig:teff_logg_rphk_cycles} 
\end{figure*}

\begin{figure*} [ht]
\includegraphics[width = 2.0\columnwidth]{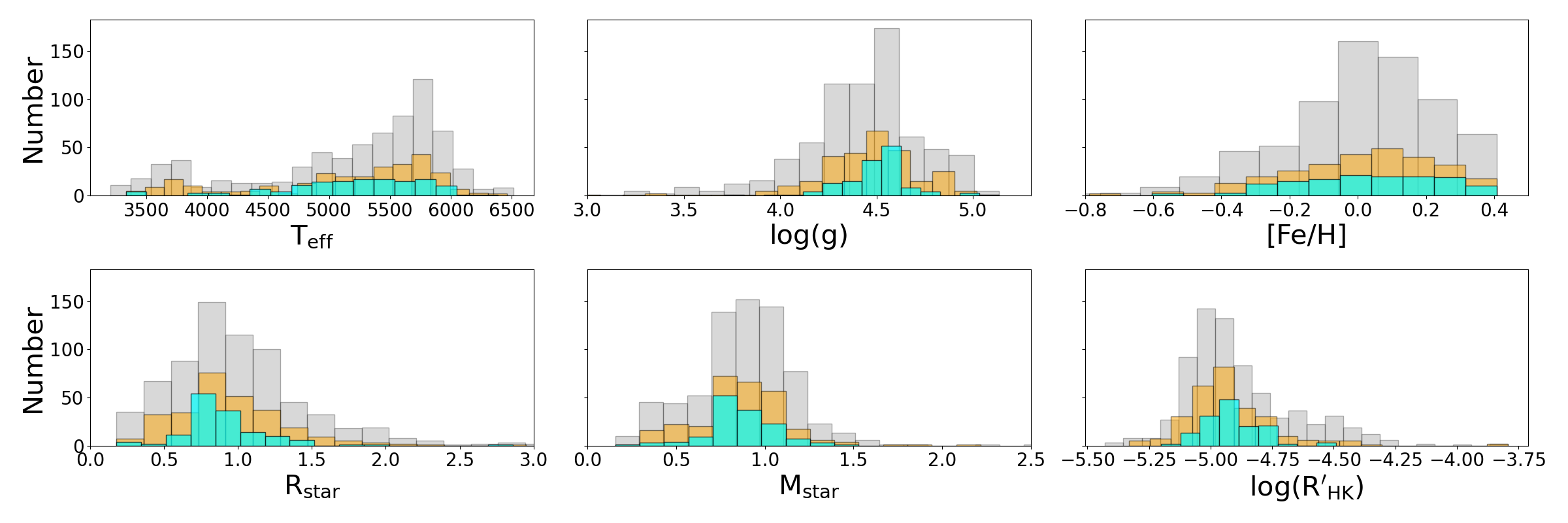}
\caption{ 
Distributions of \teff, \logg, \feh, \rstar, \mstar and \rphk for 710 stars in our sample. The yellow areas show the 256 stars that have more than 45 observations, and cyan identifies the 138 stars with a cycle from this work. The stellar parameters of the stars that were searched and those that have cycles are representative of the entire sample, and are not strictly confined to particular stellar properties. Twenty-eight stars have radii larger than 2.5.} 
\label{fig:hist_teff_logg_feh_more} 
\end{figure*} 

\vspace{0.1in} 
\section{Stars With Cycles}   
\label{sec:starswithcycles}

We examine the population of stars with identified cycles, starting within the context of fundamental stellar properties. Stars with cycles span a range of temperature (6385 K > \teff > 3332 K ), stellar surface gravity (5.13 > \logg > 3.71), and metallicity ( -1.61 > \feh 0.41). The second least metallicity is -0.56.  Figure \ref{fig:teff_logg_rphk_cycles}, similar to Figure \ref{fig:teff_logg_rphk}, identifies stars that were searched and those with stellar cycles in the \teff vs \logg parameter space. Most cycling stars are main sequence with temperatures from 4700 - 5900 K. Metallicity does not appear to hold a pivotal role in the presence of chromospherically detected activity cycles (see Section \ref{sec:cycles_metallicity}. Stars with detected cycles are listed in Table \ref{tab:table_cycles} along with their cycle properties. 

In the following section, we divide stars below 4700 K from those with higher \teff. This divide exists for two reasons. Near 4700 K, the models used to determine fundamental stellar parameters change due to the underlying physics inside stars. For example, stars below this divide use Specmatch-emprical \citep{Yee2017} and stars above use SpecMatch-synthetic \citep{Petigura2017}. The second reason is the historical division of stars at this \teff for which rotation periods and convective turnover times were devised \citep{Noyes1984}. 

Figure \ref{fig:hist_teff_logg_feh_more} shows histograms of the fundamental stellar parameters \teff, \logg, and \feh as well as the derived parameters \rstar, \mstar and \rphk. Grey identifies the 710 star sample, yellow represents the 285 star sample that we search for cycles and cyan represents detected cycles.

\begin{figure*}       

\includegraphics[width = 2.2\columnwidth,trim = {3cm 5cm 2cm 7cm},clip]{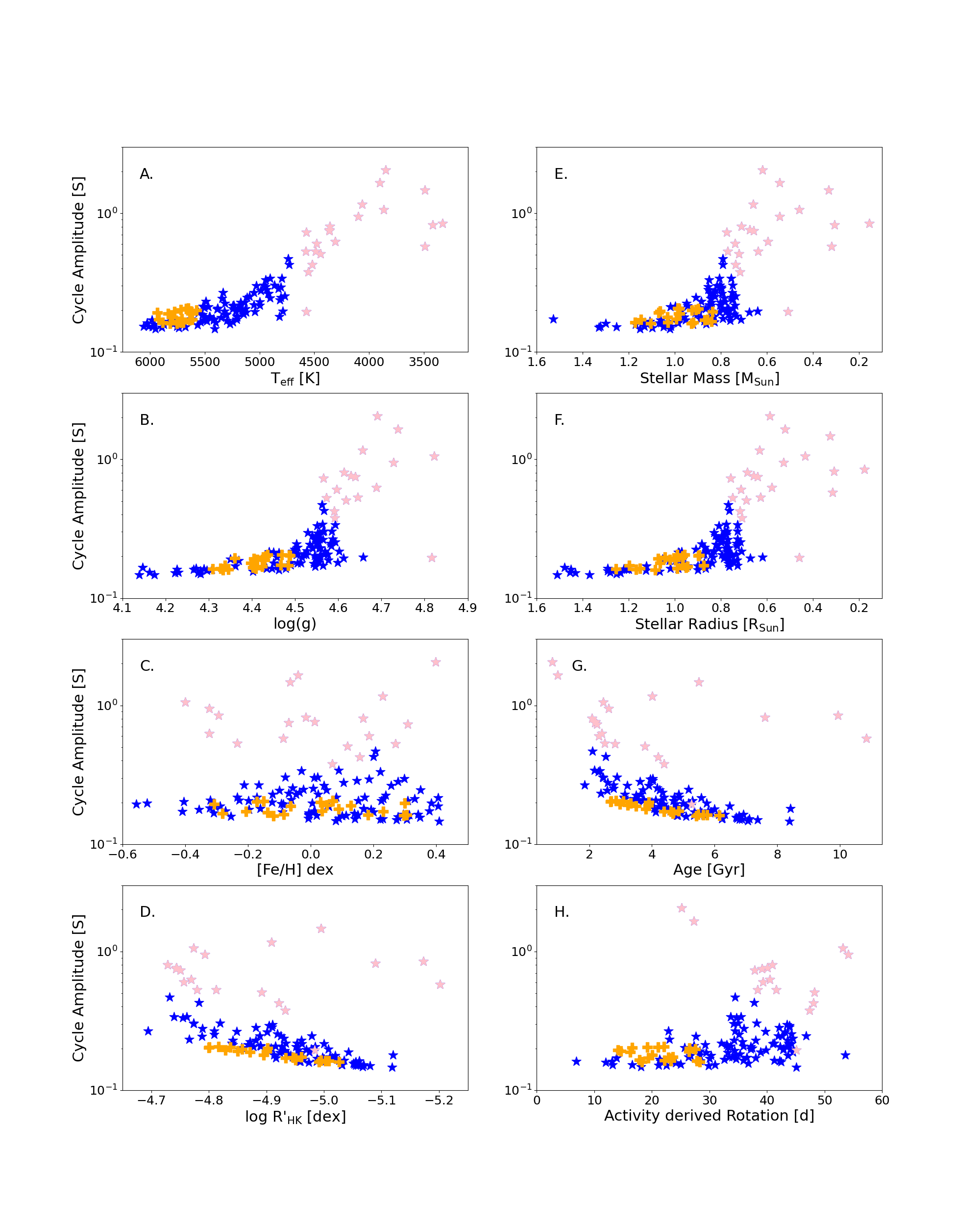}
\caption{ 
Stars with stellar activity cycles are plotted with their cycle amplitude as a function of fundamental stellar properties, \teff, \logg, \feh and \rphk (panels A, B,C, D, respectively).  The derived parameters of \mstar, \rstar, chromospheric age and activity derived rotation period are plotted in panels E, F, G, H, respectively. Age uncertainties are 60\% \citep{Mamajek2008}. Orange crosses highlight solar-type stars with 5600 K < \teff < 5900 K. Pink stars have \teff less than 4700 K. Note their distinct parameter space from the FGK stars. The ages and rotation periods for these cooler stars are not well calibrated with activity and are only shown for completeness. Cycle amplitudes are S-value peak amplitudes, not peak to peak. HD 38529 with a radius of 2.5 \rsun is not shown in panel F. }
\label{fig:cycles_amplitudes} 
\end{figure*}

\begin{figure*} [ht]  
\includegraphics[width = 2.2\columnwidth,trim = {3cm 5cm 2cm 7cm},clip]{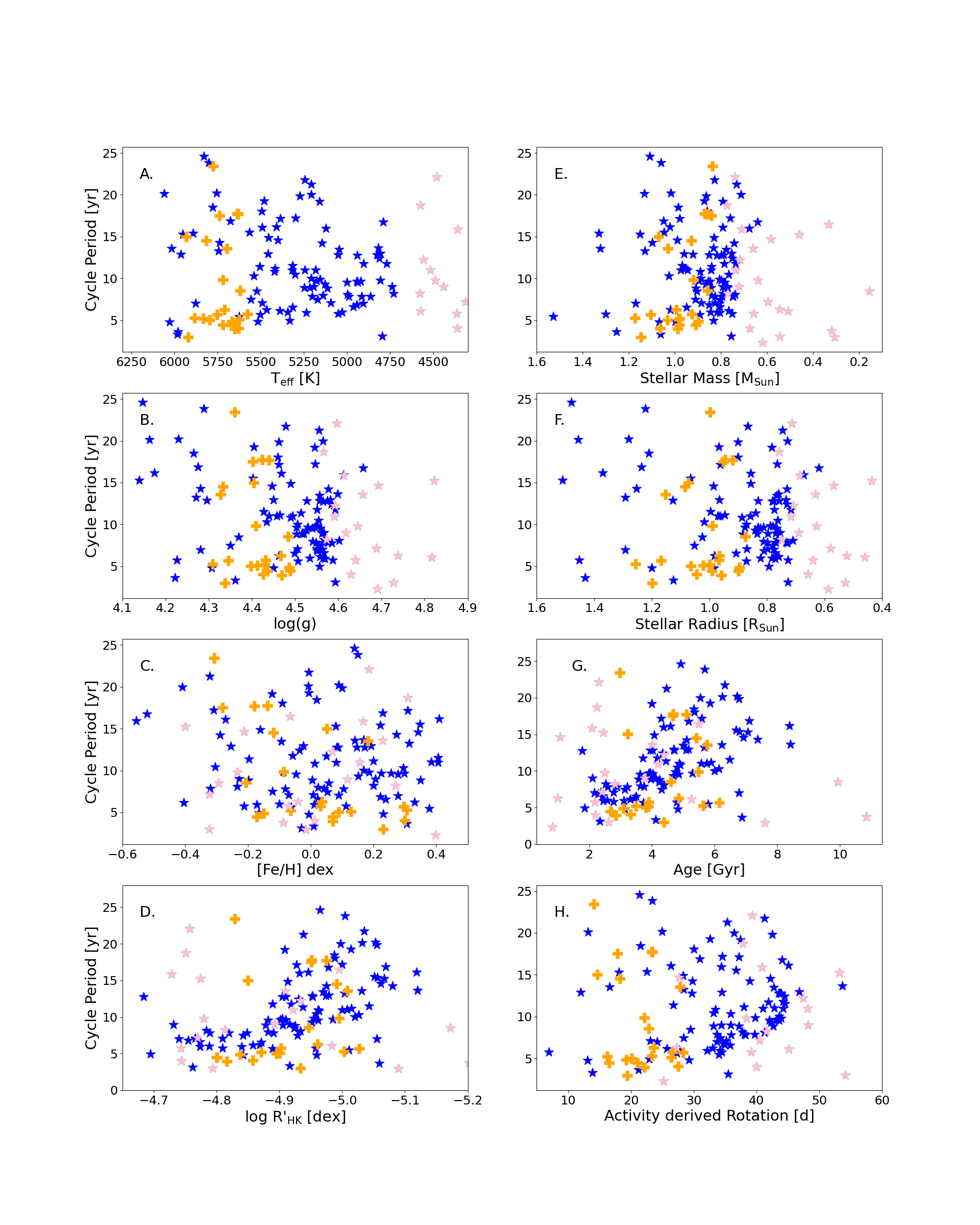}
\caption{ 
Stars with stellar activity cycles are plotted with their cycle period as a function of their fundamental stellar properties, \teff, \logg, \feh and \rphk (panels A, B, C, D, respectively) and derived parameters \mstar, \rstar, chromospheric age and chromospheric rotation period (panels E, F, G, H). Orange crosses highlight solar-type stars with 5600 K < \teff < 5900 K. Stars with \teff < 4700 K are plotted in pink. Note their distinct parameter space from the FGK stars. The ages and rotation periods for these cooler stars are not well calibrated and are only shown for completeness.}
\label{fig:cycles_periods} 
\end{figure*}

Figures \ref{fig:cycles_amplitudes} and \ref{fig:cycles_periods} show the amplitudes and periods of the detected cycles as a function of both fundamental and derived stellar properties. In amplitude, stellar cycles tend to increase as \teff decreases and main-sequence stars have larger amplitudes than evolved stars, but some main-sequence star cycle amplitudes are comparable to those of subgiant stars.  We find no correlation between metallicity and  cycle amplitude nor cycle period. A quantitative analysis is detailed in Section \ref{sec:cycles_metallicity}. We explore an intriguing correlation of cycle period to \teff for the activity range between \rphk  of -4.7 to -4.9 in Section \ref{sec:discussion}.

\subsection{Solar-type Stars with Cycles}

Previous surveys of chromospheric activity focused on solar-like stars identified with similar B - V colors and bolometric luminosities to the Sun \citep{Baliunas1995,Henry1996}, and we define our solar-similar sample to have 5600 K < \teff < 5900 K, with no metallicity nor \logg restriction. Out of 70 solar-like stars, we find 29 stars with cycles having periods between 3.9 and 23 years. This sub-sample is evenly distributed in metallicity ( $\pm 0.3$ dex) and \rphk (-4.8 to -5.08). When the average activity is used to derive the rotation period and age \citep{Mamajek2008}, the solar-sample ranges from 14-34 days and 2.7 to 7.4 Gyr, respectively. The range of cycle periods for this sub-sample offers insight into the decreasing activity and rotation as a function of age for solar type stars. The cycle periods of solar-type stars and their dependencies on stellar properties are highlighted as yellow crosses in Figures \ref{fig:cycles_amplitudes} and \ref{fig:cycles_periods}. 

\subsection{Short-Period Cycles}
We identify 45 cycling stars with periods less than 7 years. \cite{Baliunas1995} found that stars with cycles of periods shorter than 7 years have a higher false alarm probability, marking them as 'fair' or 'poor'. We use numeric thresholds that can be used to mark the robustness of a detection but find the vastly different sampling can require judgment calls when defining thresholds. HD 218868 is an example of a convincing with a period of 4.8 years.  Since the Mt. Wilson survey had sufficient sampling and sensitivity for detecting short period cycles, their sample selection is likely responsible for non-detections. Younger stars (Age less than 1 Gyr), such as HD 22046 have previously identified cycles, but we intriguingly find 15 stars near solar temperature with periods less than 7 years (Section \ref{sec:discussion}).

\subsection{Studies of Fully Convective Stars}
\label{sec:fullyconvective_found}

As with fully radiative stars, stars with masses below 0.35 \msun become fully convective and lose their tachocline, requiring a different mechanism for generating magnetic fields compared to solar-type stars \citep{Irving2023}.  Four fully-convective (\teff less than 3500K) stars have periodic activity that pass our thresholds (Section \ref{sec:cycle_search}): HD 239960, GL 273, GL 699, and HIP 74995. HD 239960 has observations during a flare that are much higher than their average activity value. We include flare stars -- except those with Helium emission that were omitted in Section \ref{sec:data_quality} -- and less active stars that we can measure with traditional S-values.  

HD 95735 has a candidate cycle showing a downward linear trend, indicating a period beyond our baseline of observations.  For fully-convective stars that are amenable to the S-value measurement, we find that the S-values are sometimes dominated by the rotation period, so our exclusion in our periodogram search below 100 days is useful. The cycles in fully-convective stars are identified with the periodogram peak method, and are not typically identified with the threshold described by Equation \ref{equation1}. 

The stellar activity of GL 699 was studied by \cite{Lubin2021}, but that study focused on periods less than 1000 days. The peak value in the GL 699 periodogram is very close to our acceptable threshold, and combining multiple datasets would provide more confidence in the detection.  A comparison between our results of GL 699 and those of \cite{ToledoPardon2019} are summarized in Section \ref{sec:gl699detect}.

\subsection{Metallicity of Stars with Cycles.} \label{sec:cycles_metallicity}
We find no correlation between cycle period or amplitude to stellar metallicity. First, we divide the stars searched into those with super-solar and sub-solar metallicity, we find similar ratios of stars searched (45\% / 55\%) to cycles found (44\% / 56\%). 
The evidence for a correlation in the flare rate of Kepler stars with metallicity found in Kepler stars \citep{See2023} does not hold for our sample. Our sample avoids stars with flares, which tend to be more active, making a direct comparison with studies of flare stars difficult.

The most metal poor star with a detected cycle,  HD 25329 has an \feh of -1.61. It has a smaller cycle amplitude compared to stars with a similar \teff by a factor of 3. The extremely low metallicity is unusual in our sample, and the CLS1 stellar parameters list the Gaia parallax as an unlikely 12 arcseconds, so the \teff may not be need revisited. It is also the outlier in panel B of Figure \ref{fig:cycles_amplitudes}, building suspicion.

To quantify the correlation between metallicity and activity, we calculate the Pearson and Spearman correlation coefficients for metallcity and a variety of activity indicators, finding no strong correlations. For correlations between \feh and cycle period, and \feh with cycle amplitude, we find coefficients below 0.10 indicating there is little or no correlation between metallicity and either cycle period or cycle amplitude. 

\subsection{Stars with Multiple Periodic Cycles}
Our time baselines are sufficient to identify stars that have multiple simultaneous cycles, but our search method did not recover any of the known occurrences. HD 22049 \cite{Metcalfe2013} of ``2.95 $\pm$ 0.03 years and 12.7 $\pm$ 0.3 years'' , HD 32147, HD 4915, HD 219234(trend), HD 4628 (trend), HD 45184 (5.14 years \citep{Flores2016}), all show evidence of a second cycle. HD 100180 shows two possible periods in \cite{Olah2016} but we only find one, and \cite{Baum2022} find none. HD 201091 and HD 201092 both have two cycles that will be apparent when combining the Mt. Wilson data. HD 18803 has one strong cycle that is changing significantly in amplitude over 20 years.  HD 219834B shows a linear trend on top of a cycle, indicative of a second cycle. The shorter of the cycles is typically not represented well by a sinusoid, so our non-detections are limited by our search method not our data quality.

\subsection{The Least Active Stars}
\label{sec:leastactive}
The search for stars in a Maunder Minimum, or magnetic-minimum state, attempts to connect the sun's activity cycles to the cycles to other stars. HD 4915 is a candidate star in a Maunder Minimum-like state \citep{Shah2018}, but our extended time baseline shows two cycles. One periodicity is at 4.9 years, and the second is more than 40 years, and the longer period cycle is now turning higher. We detect the 4.9-year period but cannot limit the period of the second cycle with Keck data alone. HD 166620 is the most convincing to have strong evidence in favor of being a Maunder minimum-like star \citep{Baum2022,Luhn2022}, and continues to show very low variation in our extended time series. 

In \cite{Isaacson2010}, the 1\% dispersion in the S-values of HD 10700 is used to gauge the systematic uncertainty of the S-values. \cite{GomesdaSilva2021} find a dispersion of 0.83\%, which is comparable to our extended time baseline for HD 10700 S-values 0.75\% (0.00125/0.1674). Our least active star  with more than 45 observations, is HD 55575 with a relative dispersion of 0.0007/0.1562 = 0.45\%. Fifty stars have a smaller S-value standard deviation than HD 10700 and should be considered the least-active, well-sampled (45 observations or more) stars in our sample (Table \ref{tab:table_average_activity}). Differentiating inactivity due to stellar evolution, stellar viewing angle and main-sequence spindown would be an interesting extension of this work. 

\subsection{Unexpectedly Cycling Stars}
For stars with stellar surface gravity less than 4.0, we find that out of 28 stars only HD 38529 (\logg = 3.93) has a detectable cycle, and its period is 6.11 years. \cite{Baliunas1995} states that ``the range of masses that can support solar-like magnetic activity is imprecisely known'', and although we expect stars to lose their cycles as they age, spin down, and evolve, this cycle is an unexpected robust detection. HD 38529 has two substellar companions, the more massive of the two has an M$\mathrm{_{sini}}$  = 13.2 M$_{Jup}$, P = 5.8 years and eccentricity = 0.35. Our periodogram analysis shows that both the peak in the RV and the S-value periodogram is at 5.8 years. This system is worthy of an analysis that explores the relationship between the planet and the activity cycle of this post-main-sequence star with stellar radius of 2.8 \rsun and a robust activity cycle. Gravitational interactions between exoplanets and stars have not been found \citep{Obridko2022}, but main-sequence stars (including our own) have been found to have activity cycles and planets with similar periods \citep{Wright2015}.

Fully radiative stars, above the Kraft Break \citep{Kraft1967}, lack a radiative-convective boundary that is known to generate magnetic fields. Without a tachocline, it is not clear what mechanism would generate magnetic activity. We find one such star with a cycle, HD 210302 (5.7 years, \teff = 6385 K). The S-value standard deviation is 0.0021 and the cycle amplitude is 0.160, making the amplitude of this cycle near the limit of detection. Most stars with \teff > 6000 K have an S-value RMS less than 0.002, the threshold below which we do not search for cycles.

\subsection{Candidate Cycles}
The choices of number of observations, and model selection drive our detection thresholds. While the well-sampled stars with high amplitude signals are straightforward to identify with numeric thresholds. Those on the margin of detection, near detection thresholds or with poor sampling  may not be considered cycling stars when additional observations are added or new detection methods are used such as the Fourier transform and the Choi-Williams distribution used in \cite{Olah2016}. 

Some candidate cycles include HD 188015 which has a strong peak in the periodogram at 11 years, but has only 38 observations. This shows that some of our declared cycles may not pass future thresholds, others will be added to the cycling star catalogs in the future.  We do not identify linear trends, which are likely indications of cycles with periods beyond our baseline.

\section{Discussion}\label{sec:discussion}
\subsection{Our Assessment}
The changing nature of a star's cycle is impossible to observe over mega-year to giga-year timescales. By collecting vignettes of hundreds of similar stars over several decades, we can piece together their long-term behavior. Observations of the solar cycle have been collected over hundreds of years, covering dozens of solar cycles. The nature of the solar cycle has been explored by observing solar-like stars' chromospheric activity on yearly and decades long time scales. The dedicated long-term observing programs that have enabled the collection of datasets that cover 20, 30 and 40 years have proven invaluable in revealing the evolution of stars stellar activity cycles.. 

High-resolution spectroscopy has contributed to these studies through time-series observations, primarily to find and characterize extra-solar planets. \cahk time series are collected alongside RV measurements to decorrelate RVs from stellar activity \cite{Mayor1995}. The California Legacy Survey provides 20 year observing baselines for 285 stars (710 on shorter baselines) on a single instrument. Planet search spectroscopy also allows determination of precise stellar properties which have been used effectively to search for subtle trends in exoplanet demographics \citep{Fulton2018}.

Using the activity time series and precise stellar properties we identify a range of stellar activity in which nearly every star is cycling. Refined B-V values that are calculated from \teff, and \feh, and are homogeneously determined, are required for identifying the \rphk range on the main-sequence for G and K-type stars in which the period of the cycle is tightly correlated to the effective temperature. In the \teff range of 4700-5900 K and the \rphk range between -4.7 and -4.9, we find cycle period increases as \teff decreases. And for stars less active than \rphk = -4.9, the correlation does not hold, and \teff is no longer closely related to cycle period.

As young, active stars with \rphk more than -4.7 spin down and expel their angular momentum, their cycles become detectable as they have a more periodic nature. Prior to reaching the steady state of cycling these stars are likely categorized as 'active/variable' in studies such as \cite{Baliunas1995} and \cite{Baum2022}. Their stellar cycles may be present but are less sinusoidal and their period is inconsistent from one cycle to the next. The activity level at which stars transition from irregular to regular cycle period is different for different \teff. For sun-like stars [5600:5900] their cycles become regularly periodic around -4.80. For the next two bins of temperature, [5300:5600], [5000:5300], and [4700,5000], the first periodic cycles are identified at -4.70, -4.76 and -4.7. The trends described in these temperature bins hold for the stars between [5900,6300], but the bin has only 10 stars. The least active star with a cycle has an \rphk = -4.85.

\begin{figure} 
\includegraphics[width = 1.0\columnwidth]{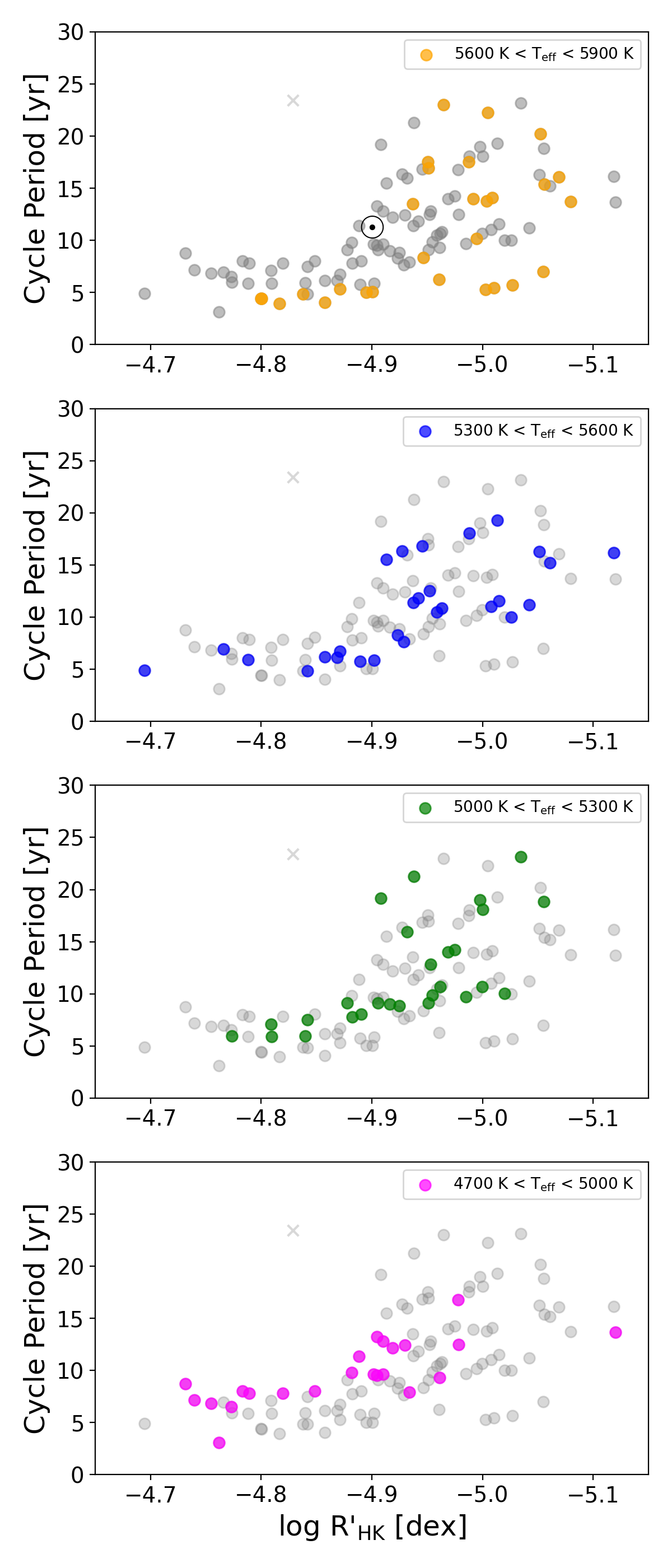}
\caption{Stellar activity cycle period is presented as a function of chromospheric activity, \rphk for different temperature ranges.  The sun is placed at \rphk of -4.9 and an 11 year cycle period. The average cycle period increases at all activity levels for every temperature bin. In the range of \rphk between -4.7 and -4.9, 34/42 stars have cycles. In each the specified range of \teff and activity, the cycle period is tightly grouped. The grey data points represent all stars between [4700,5900].}
\label{fig:cycles_periods_teff_scale} 
\end{figure} %

\begin{figure*} 
\includegraphics[width = 2.0\columnwidth]{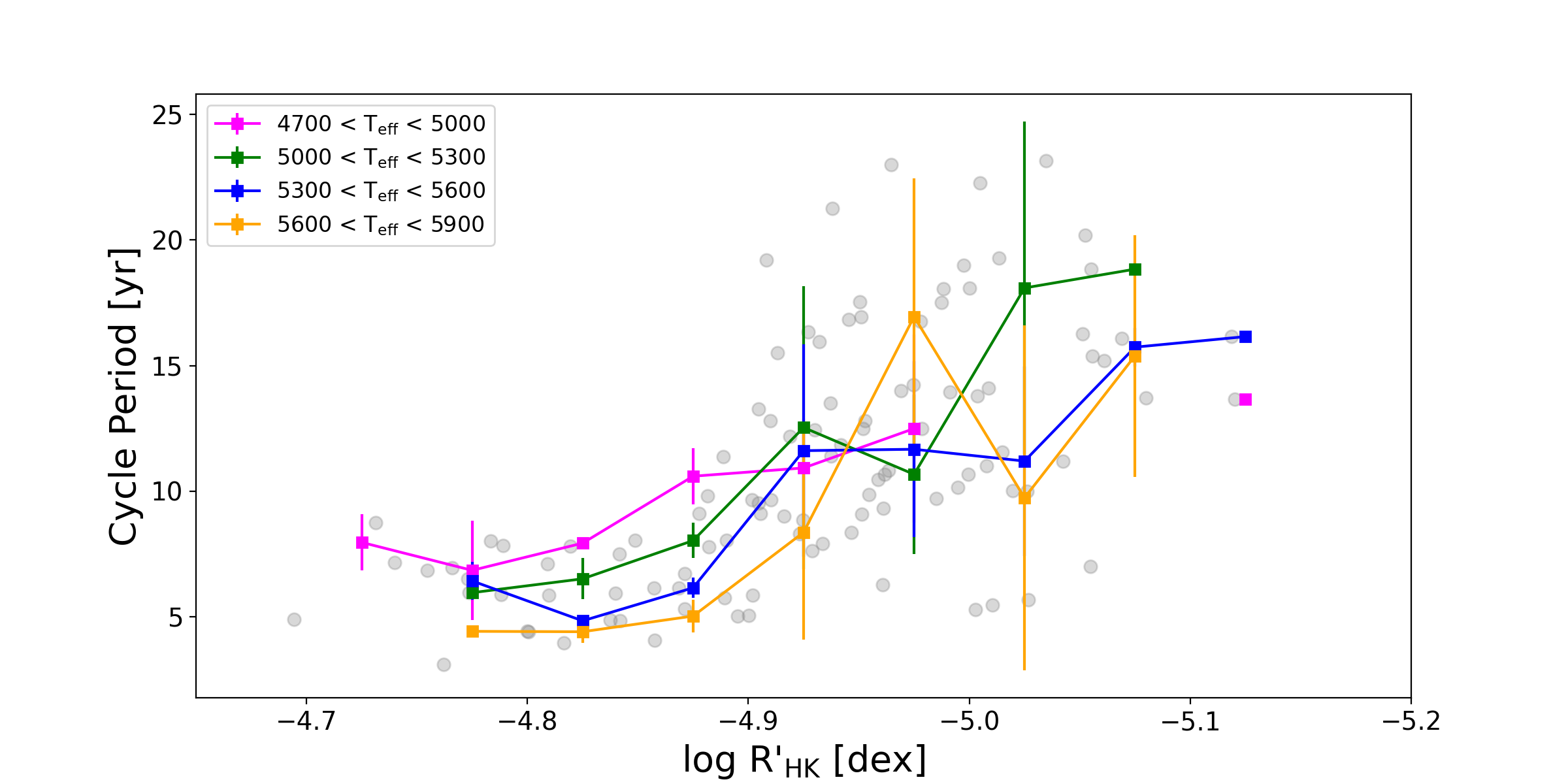}
\caption{The average cycle period and the cycle period scatter increases for every temperature bin at  \rphk = -4.90. For more active stars, cycle period and \teff are tightly coupled. For less active stars, cycle period is not related to \teff. The bin size is 0.05 and error bars represent the standard deviation in each temperature/activity bin.}
\label{fig:cycles_periods_teff_binned} 
\end{figure*} %

Figure \ref{fig:cycles_periods_teff_scale} shows the relationship between \rphk and cycle periods for ranges of temperature in 300 K bins, revealing the transition of cycle period trends at an \rphk value near -4.9. Divided at -4.9, more active stars have tightly grouped periods for each temperature range and the correlation disappears for less active stars. If we consider the stars more active than \rphk of -4.9, we find that in the solar temperature bin, cycles are 4.4 years +/- 0.5 years. From 5300-5600 K cycles are 6.0 years +/- 0.7 years. From 5000-5300 K cycles are 7.2 years +/- 1.1 years. From 4700-5000 K cycles are 7.8 years +/- 2.0 years. 

For stars less active than \rphk of -4.9, activity and cycle period decorrelate and the deterministic nature of cycle period as a function \rphk no longer holds. Cycle periods and standard deviation values are from the hottest to the coolest bin: 13.7 +- 5.6 years, 11.7+- 3.6 years, 12.8 +- 4.6 years and 12.2 +/- 2.4 years. In Figures \ref{fig:cycles_periods_teff_binned}, we average the cycle periods in bin sizes of 0.05 and plot the median with the standard deviation in each bin as the error bar, revealing the small scatter and tightly correlated cycle periods. The transition to longer periods as a function of activity occurs near \rphk = -4.9.  

Of the 42 stars in this temperature-activity range, 34 stars have confirmed cycles. The remaining 8 stars have candidate activity cycles signal that do not meet our threshold requirement or periodogram peak power. Some of them do not have the periods we expect from this newly discovered correlation. We consider this tentative evidence that every star with \teff between 4700 and 5900 K, passes through a phase in which a strongly periodic signal exists within a narrow range of periods, and this period is a function of temperature. For this to be true, we must explain why there are no cycles in these eight exceptional stars.

The stars with \teff between 4700 and 5900 K with \rphk between -4.7 and -4.9 that do not have cycles that pass our threshold are HD 159222, HD 185414, HD 176377, HD 68017, HD 37124, HD 51419, HD 212291, HD 23356, and HD 92719. HD 159222 (\teff = 5876 K) has a 3.1 year candidate cycle that passes our threshold but has a secondary peak that strikes this from our final list of cycles.  HD 185414 (\teff = 5845 K) has a candidate period at 10 years, but its periodogram peak of 0.40 falls below our threshold of 0.5. HD 176377 (\teff = 5804 K) has a candidate cycle at 4.8 years but its periodogram peak of 0.41 falls below our threshold. If future observations confirm this cycle, the cycle period would be consistent with our trend. HD 68017 (\teff = 5712 K) has a candidate cycle at 1.1 years and is slightly less active both in median S-value and S-value standard deviation. This star is potentially slightly more evolved than the others. A 1.1 year cycle would not fit our trend. HD 37124 (\teff = 5698 K) has a potential cycle 22.6 years, but the cycle is not closed and the periodogram peak is ambiguous to higher periods, so we consider this a lower limit. It also has a slightly lower \logg than the other stars discussed here. HD 51419 (\teff = 5775 K), marked with an X in Figure \ref{fig:cycles_periods_teff_scale} is very similar to HD 37124 in that the cycle is not closed and the peak at 23.4 years is not unique, providing only a lower limit on the period. Its \logg value is 4.36. HD 212291 (\teff 5589 K) has a two strong peaks in the periodogram, at 4.9 and 5.6 years, but is not uniquely determined. Further observations would likely confirm the period and it would fall into our expected trend. HD 23356 (\teff =4976 K) has a candidate signal at 5.4 years, but is only identified after removing a linear trend, and even then it does not pass our periodogram peak threshold. If this cycle were confirmed, it would fit our trend. HD 92719 (\teff =  5774 K) has a  cycle at 4.6 years, but has an ambiguous period, with a second periodogram peak, removing it from our list. 

For those stars in this exception list that do not have candidate cycles, which could be confirmed with more observations, possible explanations include an undervalued \teff, which could shift the star into the \rphk range where we do not expect a cycle. This possibility is supported by the notion that the coolest star without an expected cycle has \teff of 5589 K, meaning every star below this value has a cycle within \rphk of -4.7 to -4.9. Another possibility is a pole-on orientation for these stars.

One star,  HD 130992 (\teff = 4796 K) has an cycle of period 3.1 years, going against our trend. We find the poor sampling of this star contributes to its potentially false detection, but it passes all of our numerical thresholds so we include it in our table.

\begin{figure*}       
\includegraphics[width = 2.0\columnwidth ]
{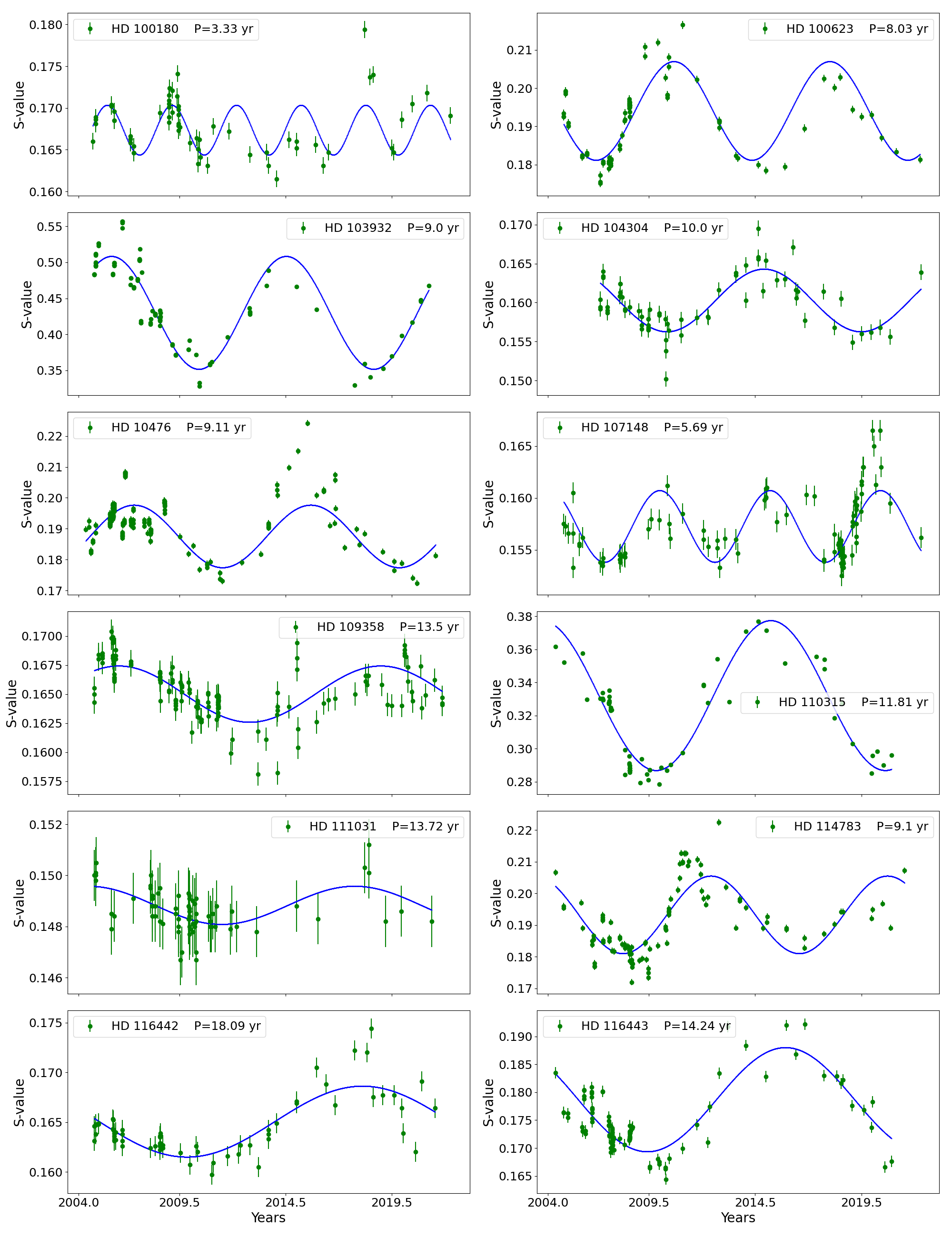}
\caption{ Cycles for stars: HD 100180, HD 100623, HD 103932, HD 104304, HD 10476, HD 107148, HD 109358, HD 110315, HD 111031, HD 114783, HD 116442, HD 116443.}
\label{fig:cycles_v1} 
\end{figure*}

\begin{figure*}       
\includegraphics[width = 2.0\columnwidth]
{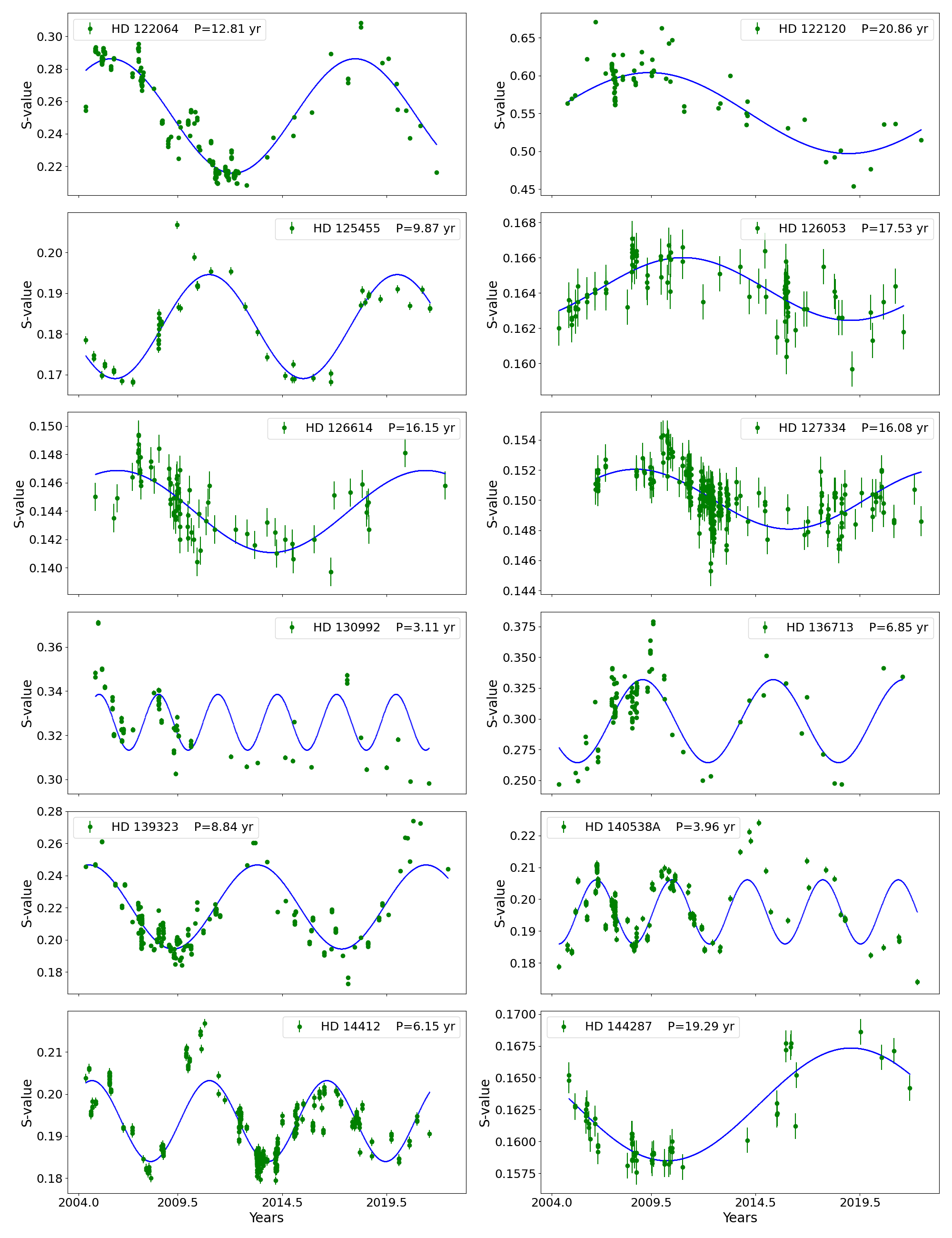}
\caption{Cycles for stars: HD 122064, HD 122120, HD 125455, HD 126053, HD 126614, HD 127334, HD 130992, HD 136713, HD 139323, HD 140538A, HD 14412, HD 144287. }
\label{fig:cycles_v2} 
\end{figure*}

\begin{figure*}       
\includegraphics[width = 2.0\columnwidth]
{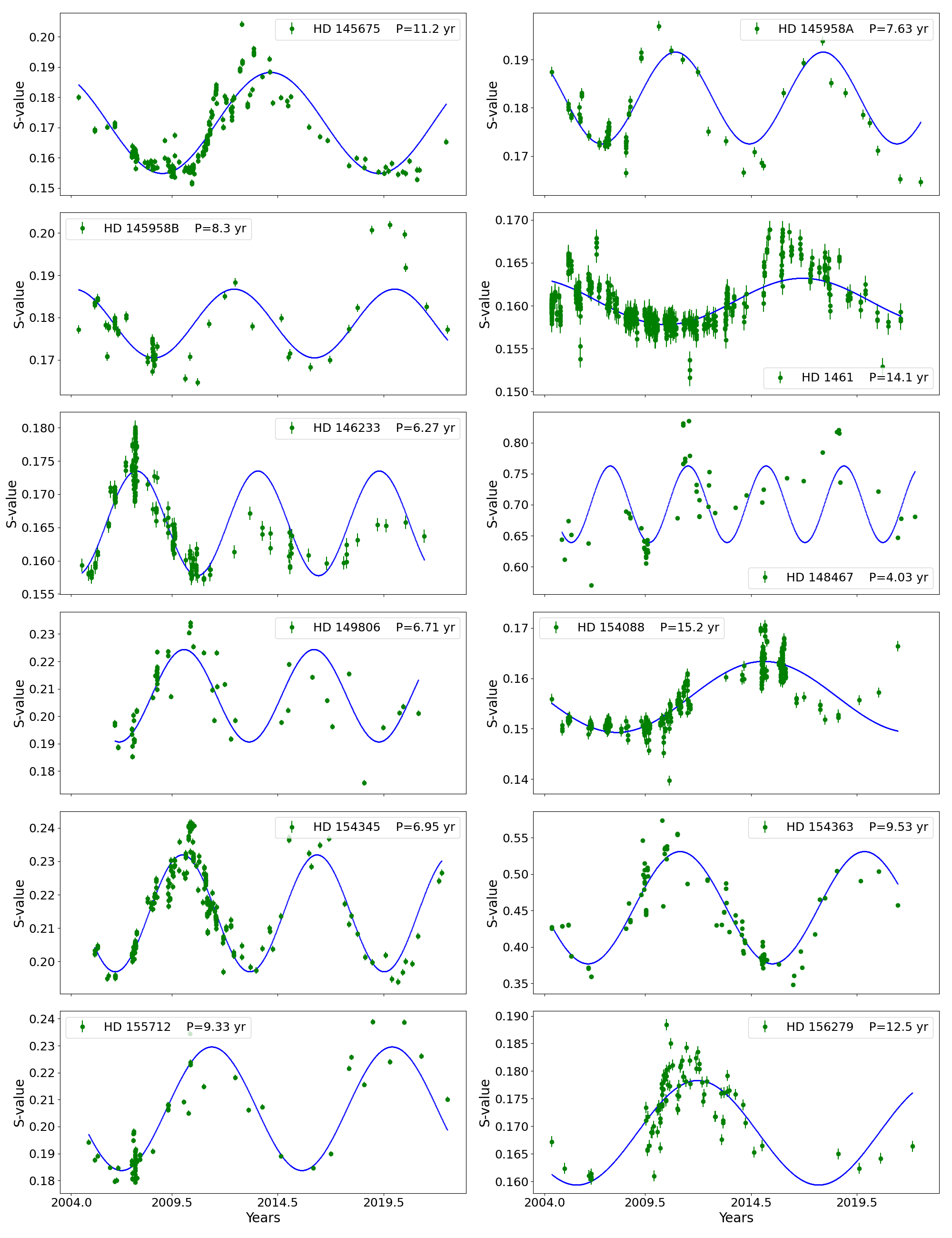}
\caption{Cycles for stars: HD 145675, HD 145958A, HD 145958B, HD 1461, HD 146233, HD 148467, HD 149806, HD 154088, HD 154345, HD 154363, HD 155712, HD 156279.}
\label{fig:cycles_v3} 
\end{figure*}

\begin{figure*}       
\includegraphics[width = 2.0\columnwidth]
{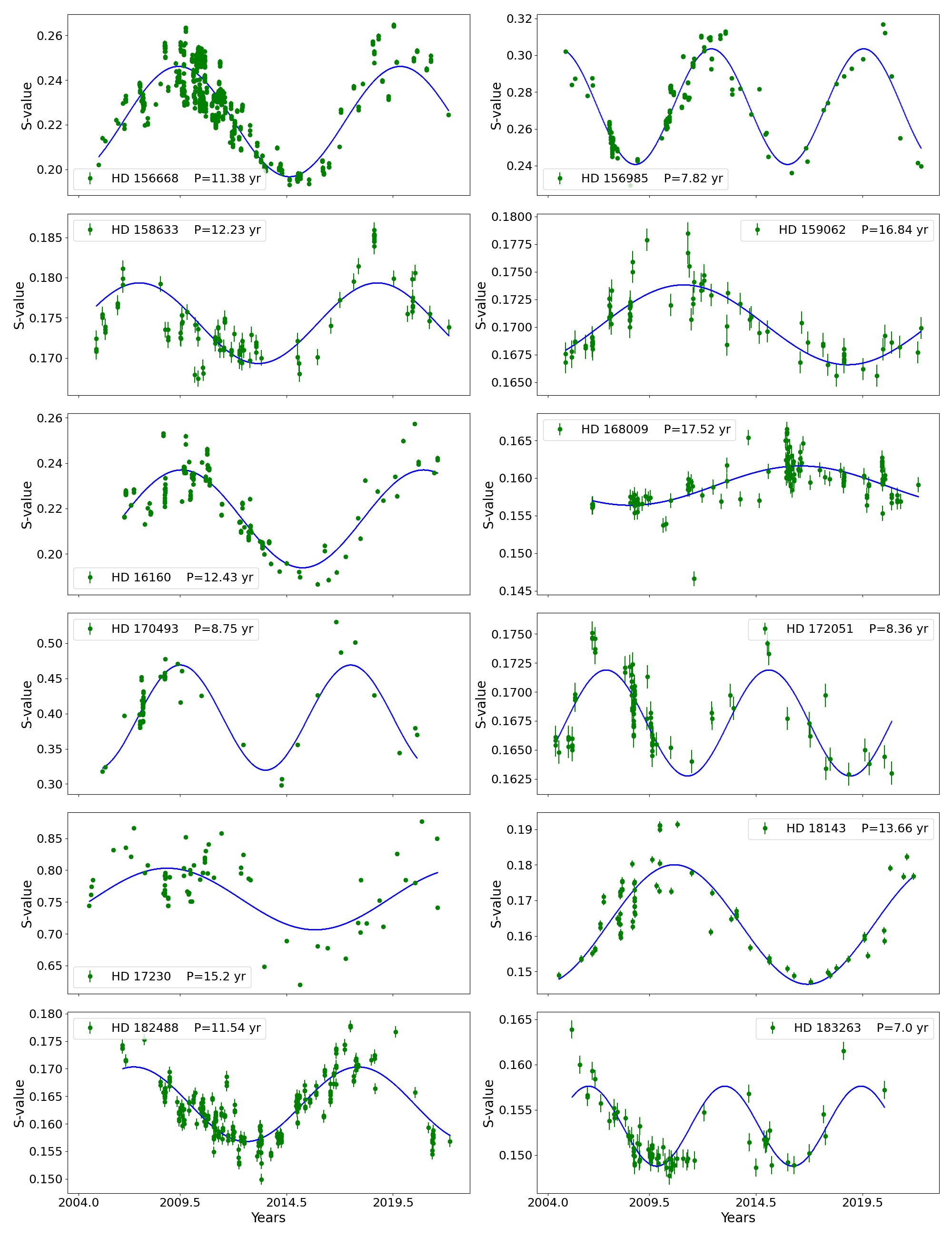}
\caption{ Cycles for stars: HD 156668, HD 156985, HD 158633, HD 159062, HD 16160, HD 168009, HD 170493, HD 172051, HD 17230, HD 18143, HD 182488, HD 183263.}
\label{fig:cycles_v4} 
\end{figure*}

\begin{figure*}       
\includegraphics[width = 2.0\columnwidth ]{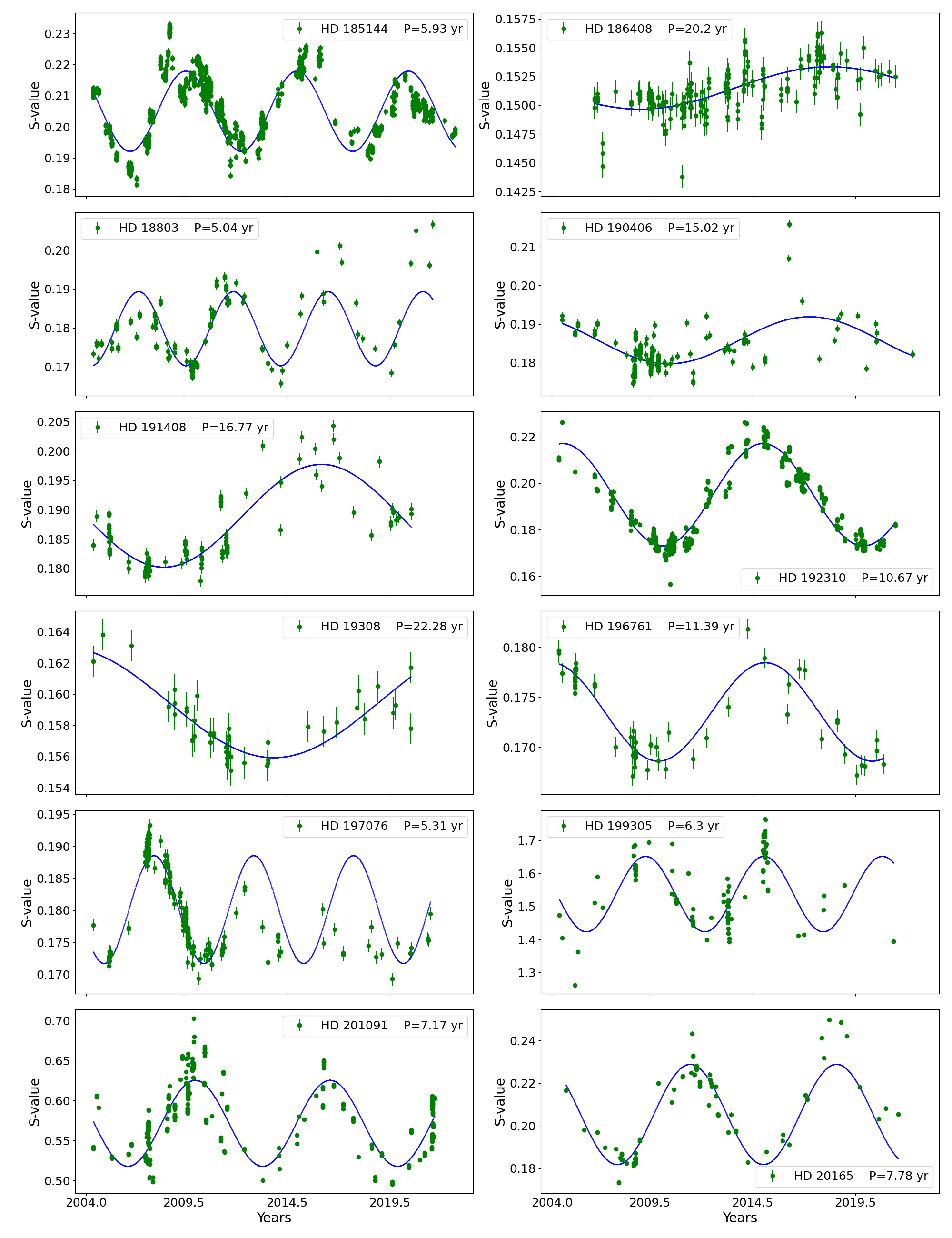}
\caption{ Cycles for stars: HD 185144, HD 186408, HD 18803, HD 190406, HD 191408, HD 192310, HD 19308, HD 196761, HD 197076, HD 199305, HD 201091, HD 20165. }
\label{fig:cycles_v5} 
\end{figure*}

\begin{figure*}       
\includegraphics[width = 2.0\columnwidth]
{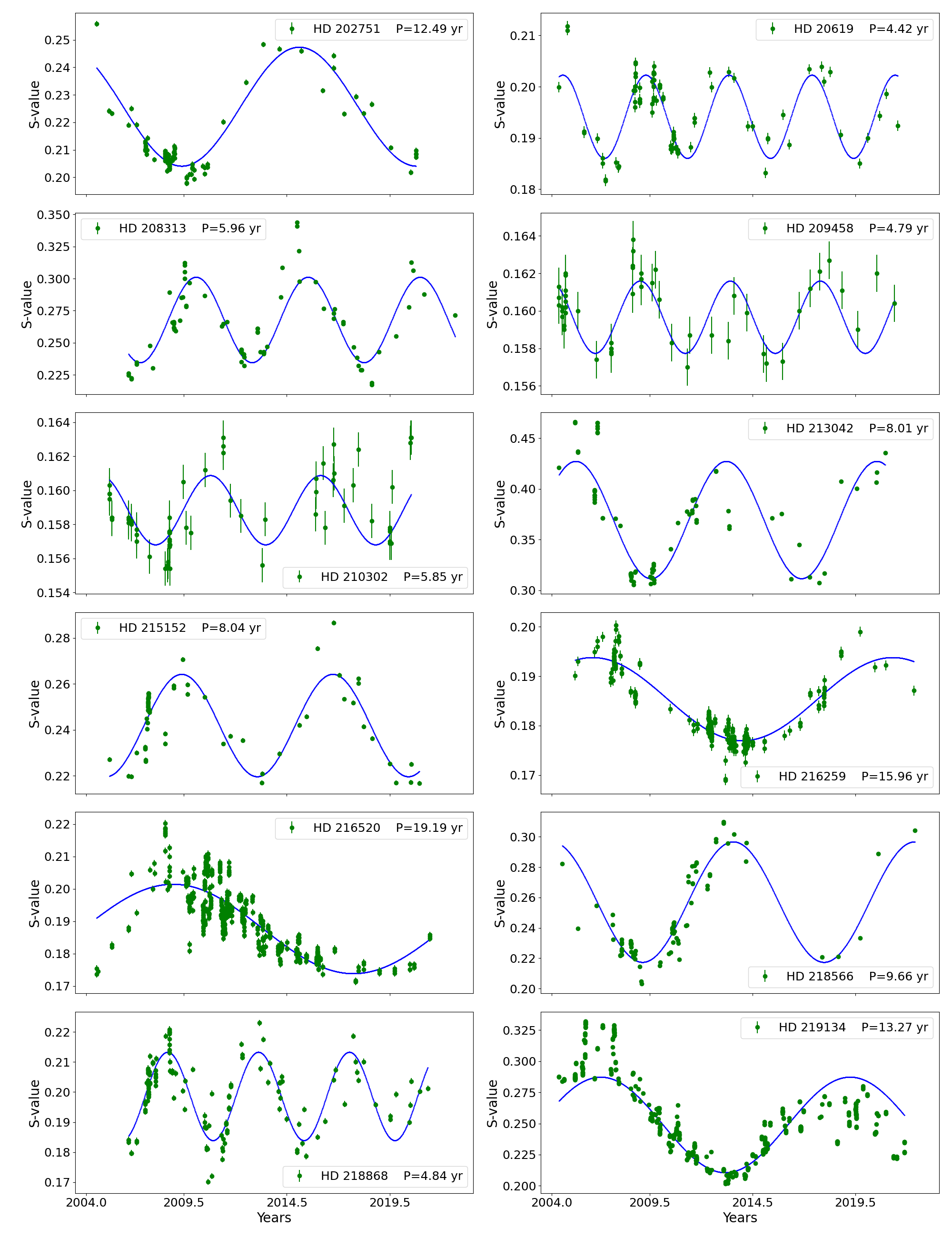}
\caption{ Cycles for stars: HD 202751, HD 20619, HD 208313, HD 209458, HD 210302, HD 213042, HD 215152, HD 216259, HD 216520, HD 218566, HD 218868, HD 219134. }
\label{fig:cycles_v6} 
\end{figure*}

\begin{figure*}       
\includegraphics[width = 2.0\columnwidth]
{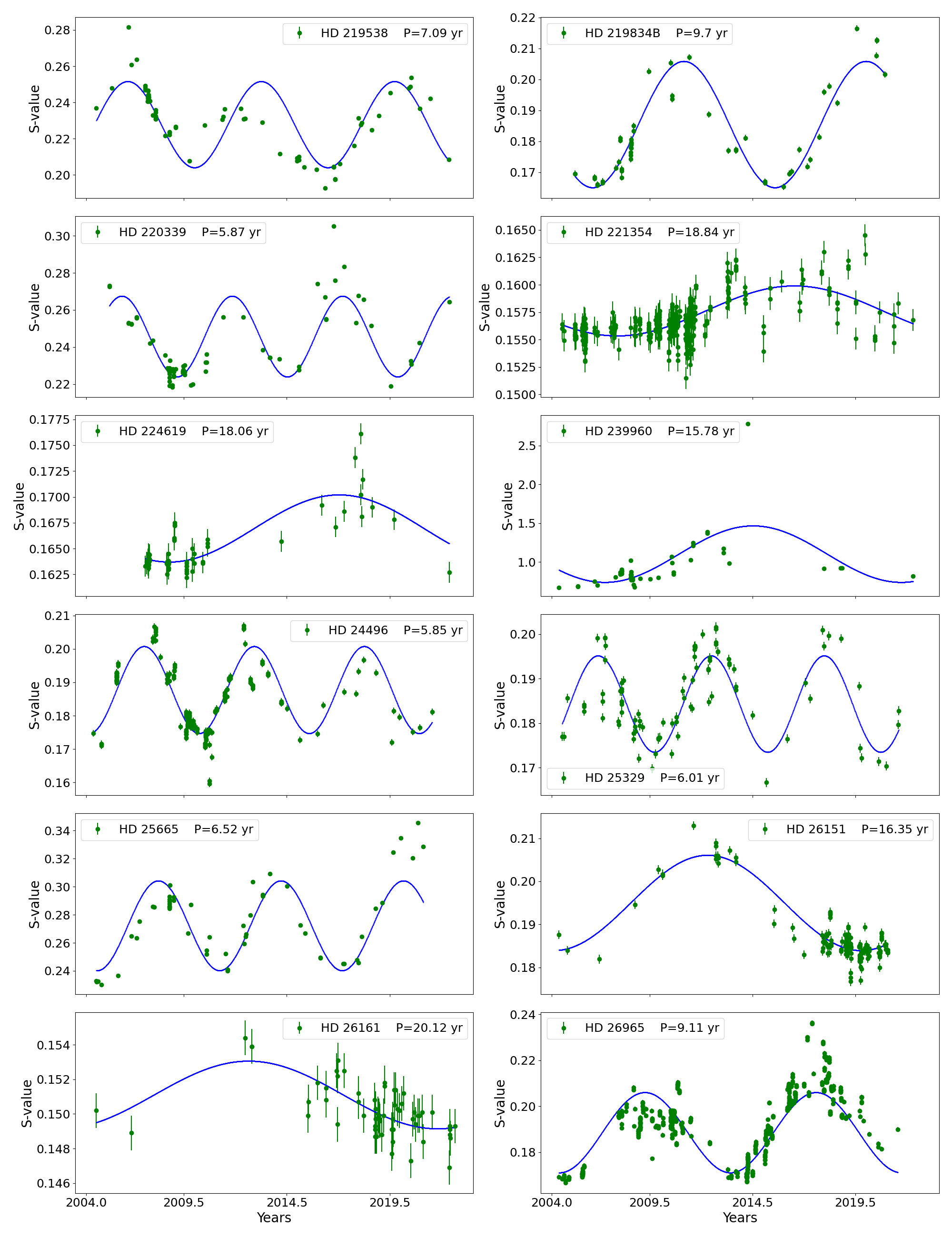}
\caption{ Cycles for stars: HD 219538, HD 219834B, HD 220339, HD 221354, HD 224619, HD 239960, HD 24496, HD 25329, HD 25665, HD 26151, HD 26161, HD 26965. }
\label{fig:cycles_v7} 
\end{figure*}

\begin{figure*}       
\includegraphics[width = 2.0\columnwidth] 
{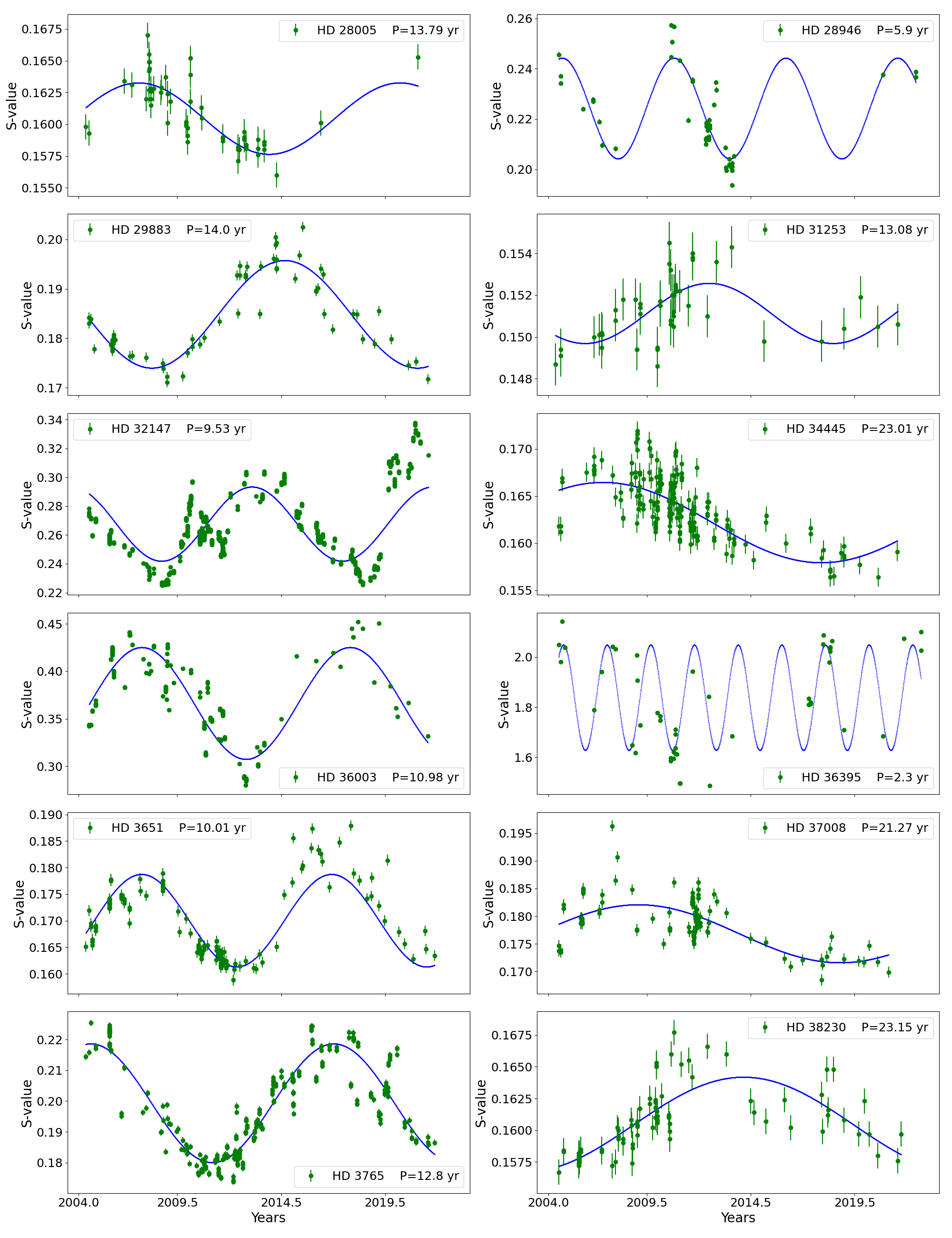}
\caption{Cycles for stars:  HD 28005, HD 28946, HD 29883, HD 31253, HD 32147, HD 34445, HD 36003, HD 36395, HD 3651, HD 37008, HD 3765, HD 38230.}
\label{fig:cycles_v8} 
\end{figure*}

\begin{figure*}       
\includegraphics[width = 2.0\columnwidth] 
{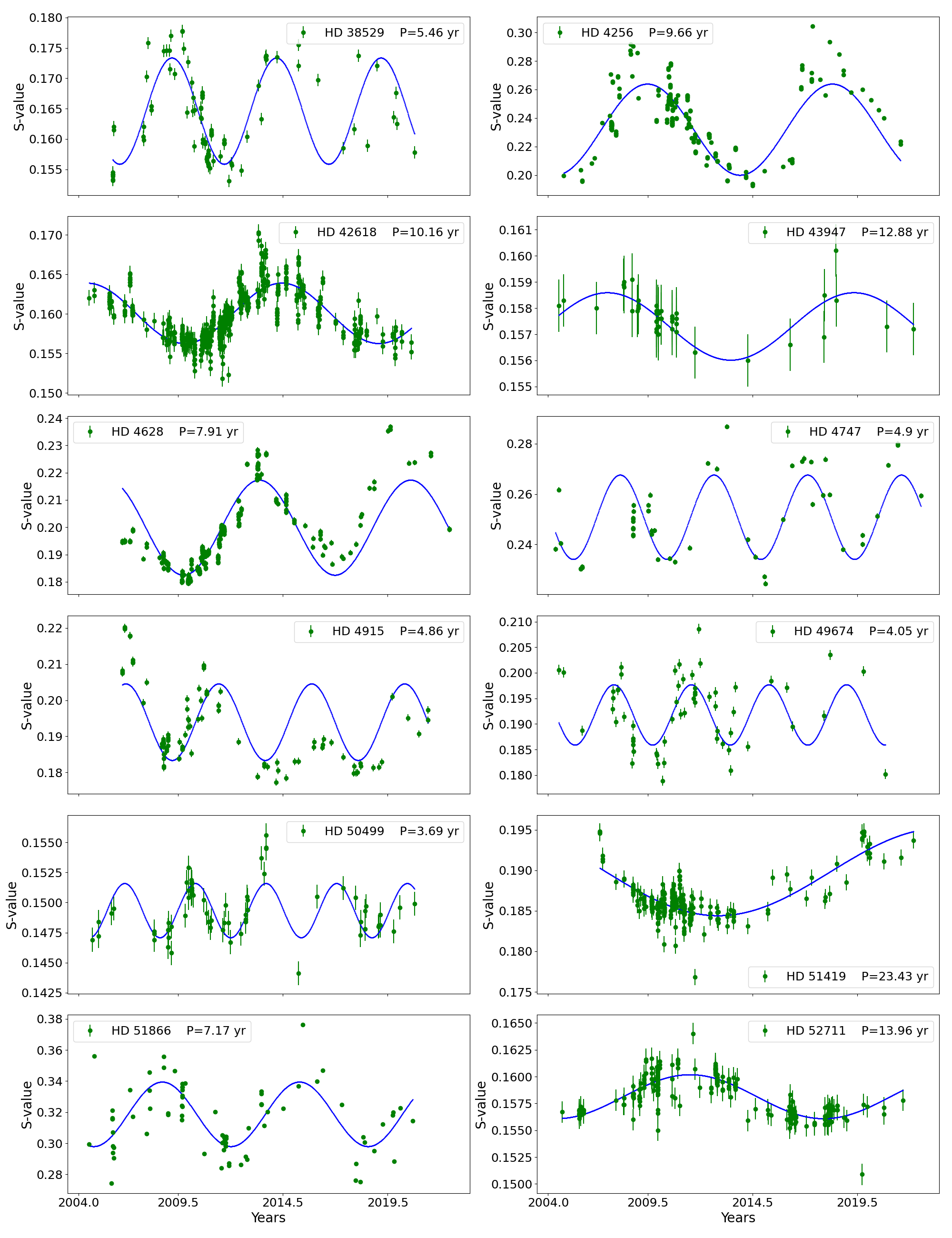}
\caption{ Cycles for stars: HD 38529, HD 4256, HD 42618, HD 43947, HD 4628, HD 4747, HD 4915, HD 49674, HD 50499, HD 51419, HD 51866, HD 52711.}
\label{fig:cycles_v9} 
\end{figure*}

\begin{figure*}       
\includegraphics[width = 2.0\columnwidth]
{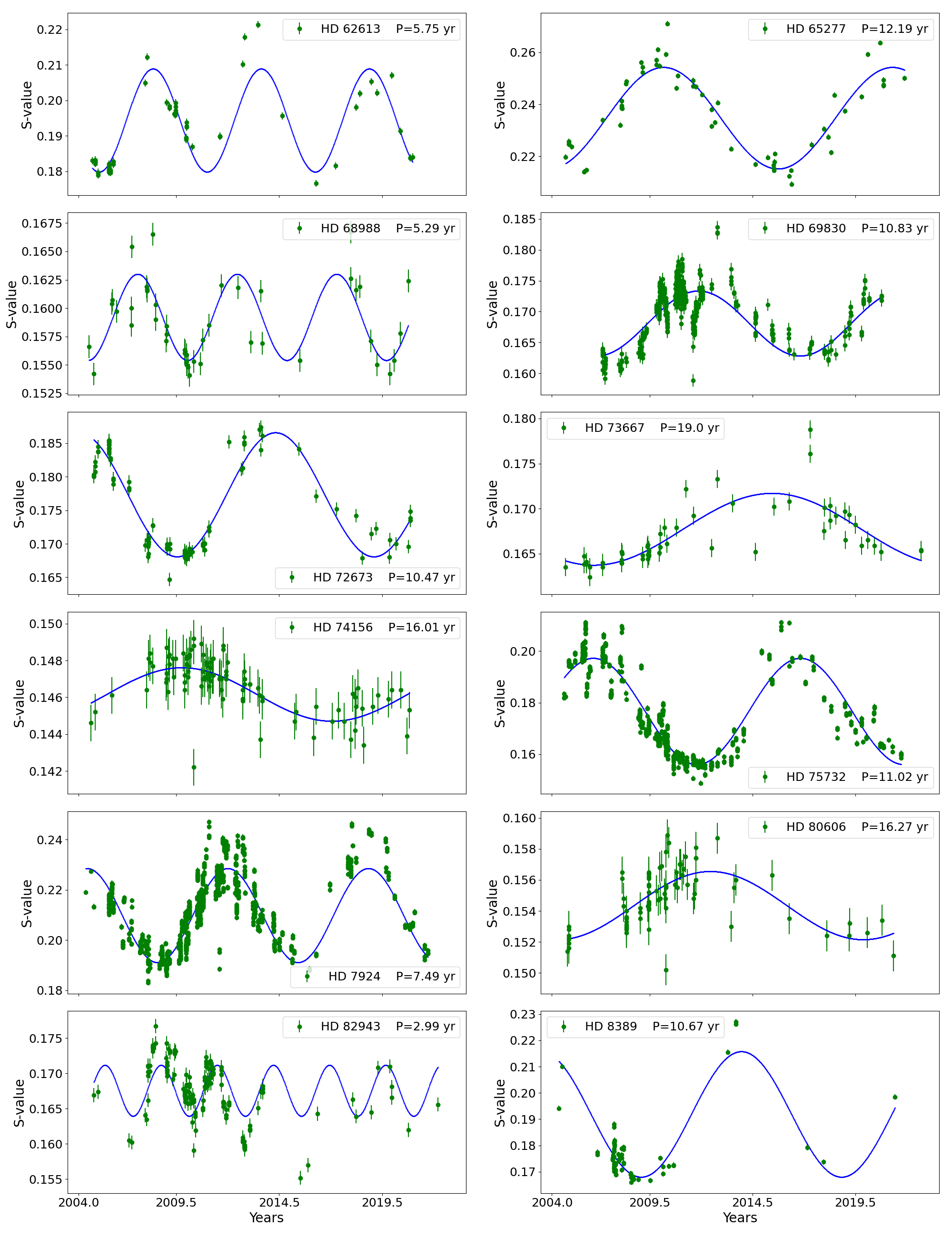}
\caption{ Cycles for stars: HD 62613, HD 65277, HD 68988, HD 69830, HD 72673, HD 73667, HD 74156, HD 75732, HD 7924, HD 80606, HD 82943, HD 8389.}
\label{fig:cycles_v10} 
\end{figure*}

\begin{figure*}       
\includegraphics[width = 2.0\columnwidth]
{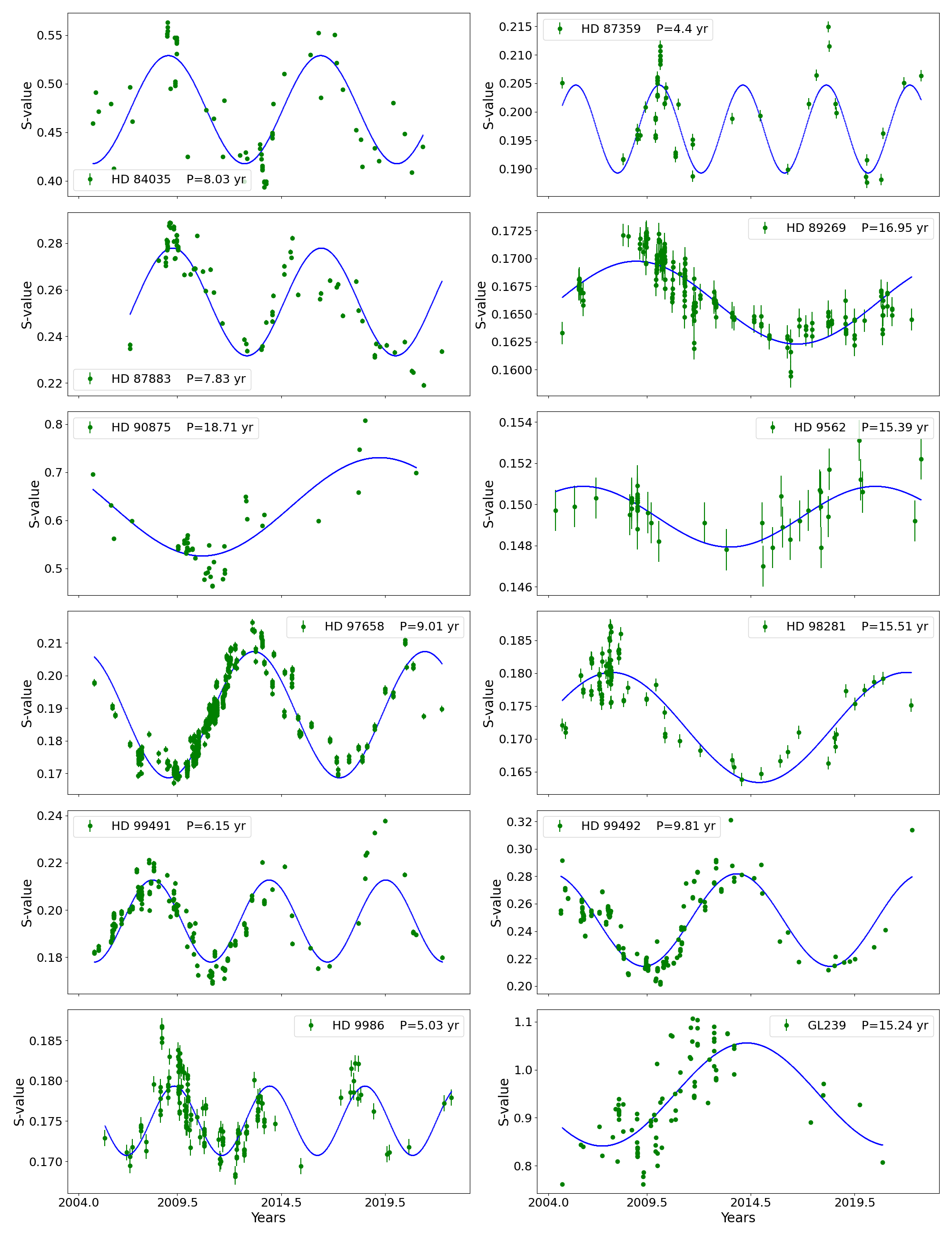}
\caption{Cycles for stars: HD 84035, HD 87359, HD 87883, HD 89269, HD 90875, HD 9562, HD 97658, HD 98281, HD 99491, HD 99492, HD 9986, GL239. }
\label{fig:cycles_v11} 
\end{figure*}

\begin{figure*}       
\includegraphics[width = 2.0\columnwidth]
{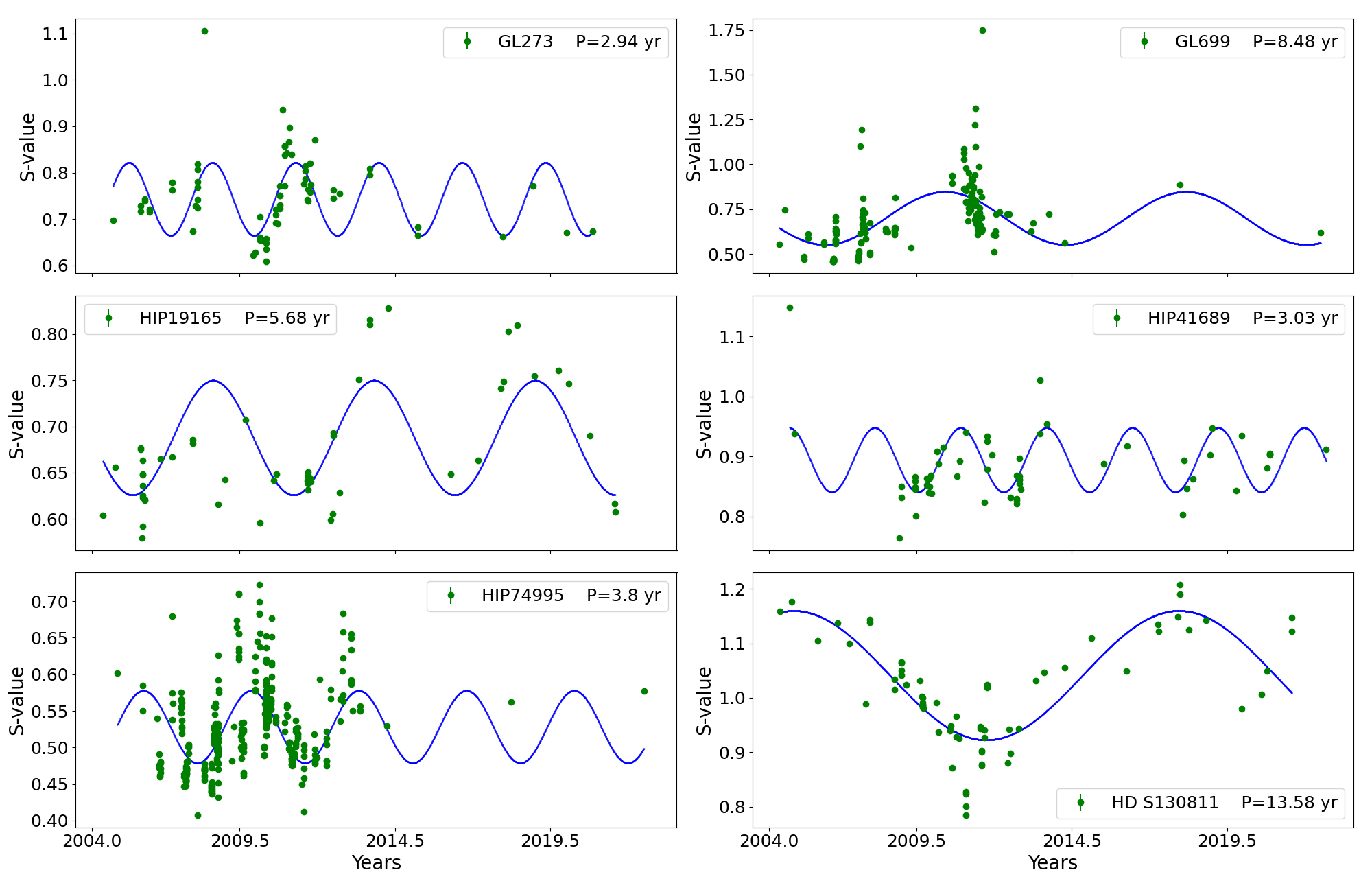}
\caption{ Cycles for stars: GL273, GL699, HIP19165, HIP41689, HIP74995, S130811.}
\label{fig:cycles_v12} 
\end{figure*}

With the possible explanations for why these eight stars do not have cycles, we again pose the possibility that every stars from \teff 4700 to 5900 K and \rphk between -4.7 and -4.9 has a regularly periodic activity cycle with a period correlated to \teff. Each of 138 stars with cycles identified in this work are presented in Figures \ref{fig:cycles_v1}, \ref{fig:cycles_v2}, \ref{fig:cycles_v3}, \ref{fig:cycles_v4}, \ref{fig:cycles_v5}, \ref{fig:cycles_v6}, \ref{fig:cycles_v7}, \ref{fig:cycles_v8}, \ref{fig:cycles_v9}, \ref{fig:cycles_v10}, \ref{fig:cycles_v11}, \ref{fig:cycles_v12}. 

\subsection{The Path Forward}

Our collection of magnetic activity cycles, found via multi-decade ground based monitoring of stars in the solar neighborhood sets the stage for further studies of magnetic activity, rotation, and age.

When considering this specific temperature range, no restrictions are placed on \logg nor \feh. Stellar evolution becomes a factor only after activity values decreases beyond \rphk of -4.9, near the Sun's activity level, at which point the changes we observe in activity cycle period become a combination of main-sequence activity changes and evolution of stars off of the main-sequence. Spectropolarimetry of solar-type stars with different Rossby numbers \citep{Metcalfe2023,Metcalfe2024} are shown to support the theory of Weakened Magnetic Breaking. Adding the findings presented in this work may add to our understanding of the Sun's activity cycle relative to other solar-type stars.

Many stars have previously noted double periods and the ratio of these periods is a strong function of \teff. The Keck/HIRES time baseline of 20 years is sensitive to cycles with a period of 25 years, but identifying a second cycle per star will require a different method and threshold of detection. When analyzing stars with two cycles, most previously studies have relied on the Rossby number, the ratio of the rotation period to the convective turn over time \citep{Mittag2023}. In this work, we notably have identified this range of consistent, predictable stellar cycle period without knowing the stellar rotation periods. 

Perhaps the most intriguing question around stellar activity cycles and rotation periods is what happens to solar type stars as their dynamo transitions from having a strong relationship between rotation, age, and activity. For old main-sequence stars, there is a breakdown between the rotation period and stellar age, but perhaps not with the overall chromospheric activity and the age of the star. The evolution of stars off of the main-sequence also clouds the interpretation of these relationships. The homogeneously determined stellar parameters from CLS1 were used to disentangle such effects \citep{David2022}. The theory of Weakened Magnetic Braking \citep{VanSaders2016,Metcalfe2022} is supported by the measurements of stellar rotation periods with \emph{Kepler} photometry and independently with asteroseismically determined rotation periods. We present another independent data set that can be used to test Weakened Magnetic Breaking.  

The examination of magnetic cycles as a function of age  \citep{Olah2016} provides a path forward for future studies that can take advantage of large time series of \cahk activity measurements. Figure \ref{fig:cycles_periods} shows that the regularly cycling stars correspond to chromospheric ages between 2 and 4 Gyr. Using independently determined ages make this conjecture more reliability. Adding measurements of rotational modulation and age to the data presented here will further elucidate the relationship between magnetic activity of stars and their observable proxies. Examination of the ratio of rotation period to activity cycle for the Mt. Wilson sample shows both consistency with previous studies and subtlety in that dependence on stellar temperature \citep{Mittag2023}. The larger sample of cycles presented here, with precise stellar properties, provides an opportunity to further study these relationships and their impact on stellar dynamos.

This new collection of stellar activity cycles, with its broad span in stellar \teff and \logg, can be used to broaden the connections between stellar cycle periods with theoretical understandings of the generation of magnetic fields in stars. 
We defer the analysis of ages, rotation periods and Rossby numbers to future studies, noting specifically that the Rossby number is not required in our current analysis. We identify the trend between cycle period and \teff and the transition from a strongly correlated period to a weak correlation with only activity time series.


\section{Conclusion}
We present the largest sample of spectroscopically determined stellar activity cycles to date, with optical spectroscopy of 710 solar neighborhood stars collected over two decades to catalog chromospheric activity, and search for stellar activity cycles. The California Legacy Survey stars forming the basis of this survey may also aid exoplanet RV surveys.  The \cahk time series data serves as a proxy for the stellar and chromospheric activity, measurements that can be utilized in the detection and characterization of exoplanets. 

From our Keck/HIRES \cahk data set, a total of 285 stars are amenable to searches for stellar cycles with periods ranging from 2 to 25 years, and 138 stars show stellar cycles of varying length and amplitude. These activity cycles observations in turn may be used to disentangle the effect of stellar magnetic activity when detecting and characterizing exoplanets. 

The results presented may also find use in placing the Sun’s stellar magnetic activity into context of the activity of solar neighborhood stars, including an improved understanding of stellar activity through a star's main-sequence lifetime.

The collection of \cahk measurements from the Mt. Wilson H and K Project helped to place the solar cycle into context in the solar neighborhood. The empirical identification of cycles and rotation periods, along with theoretical underpinnings of convective turnover times and mixing lengths has greatly improved the understanding of magnetic phenomena on and below the stellar surface. Folding stellar age into what we know about activity cycles and rotation may lead to deeper understanding of the changes in stars' chromospheric activity on Gigayear timescales. 

Finally, we provide tentative evidence that every G and K type star passes through a stage of stellar activity in which stellar activity cycles are present and their period is strongly correlated to their effective temperature.

\section{Acknowledgments}

We thank Daniel Huber, Simon Albrecht, Simon Murphy, and Aaron Householder for helpful comments, and Lee Rosenthal for his early work on CLS. And we thank the anonymous referee for constructive feedback during review.

This work was supported by NASA Keck PI Data Awards, administered by the NASA Exoplanet Science Institute. Some of the data presented herein were obtained at the W. M. Keck Observatory, which is operated as a scientific partnership among the California Institute of Technology, the University of California and the National Aeronautics and Space Administration. The Observatory was made possible by the generous financial support of the W. M. Keck Foundation. This research made use of NASA's Astrophysics Data System.

This project would not have been possible without major allocations of Keck telescope time from the University of California, California Institute of Technology, the University of Hawaii, and NASA. This work utilized the SIMBAD astronomical database.

Author Contributions: H.I. Conducted the analysis and wrote the paper. A.W.H. was an originator of the survey. B.F., E.A.P., L.M.W, are original collaborators on the CLS survey. S.K. and B.C. advised on the analysis. Authors C.B to N.S. provided comments on the paper and contributed to the observing effort in order of their appearance.

Data collected here, previously published and novel, was gathered on over 1500 individual nights by 152 unique observers. Without their contribution to astronomical data collection, this work would not be possible.

The authors wish to recognize and acknowledge the very significant cultural role and reverence that the summit of Maunakea has always had within the indigenous Hawaiian community. We are most fortunate to have the opportunity to conduct observations from this mountain.


\vspace{5mm}
\facilities{Keck I (HIRES)}


\software{We made use of the following publicly available Python modules: Astropy \citep{astropy:2013, astropy:2018, astropy:2022}, matplotlib \citep{Hunter2007}, numpy \citep{vanderwalt2011}, scipy \citep{2020SciPy-NMeth} and pandas \citep{mckinney-proc-scipy-2010}. Interactive Data Language (IDL) was used to extract the spectral line information (ENVI version 4.8 (Exelis Visual Information Solutions, Boulder, Colorado). 
}

\begin{table*}
\centering
\scriptsize
\caption{Gaussian Fit Parameters for the Activity Sample \label{tab:table_fits}}
\begin{tabular}{lrrrrrrrr}
\hline
      Property Bin & Amplitude 1 & Mean 1 & Sigma 1  & Amplitude 2 & Mean 2 & Sigma 2 & Chi-squared & Reduced Chi-Squared \\

\hline
 Full Sample (710)   & 100.2  & -4.990 &  0.1055 & 22.25   & -4.632 & 0.2136 & 495      & 20.6 \\
 Comparison (564)    & 48.58 & -4.985 &  0.097 & 6.81     & -4.599 & 0.24  & 490     & 14.4 \\
 F-type stars (47)   & 7.757 &-5.018 &  0.0508 & 2.764    & -4.833 & 0.178 & 19.7      & 2.19 \\
 G-type stars (301)  & 35.98 & -4.985 &  0.094  & 2.042    & -4.483 & 0.272  & 280     & 11.7 \\
 K-type Stars (171)  & 8.216 & -4.950 &  0.068  & 4.617     & -4.734 & 0.117 &  190&    6.13   \\
 K-type Stars (Con't)& 6.18 & -4.443 &   0.044 &...        &...    &...    &   ... & ... \\ 
\hline
\end{tabular}
\end{table*}

\begin{table*}
\centering
\scriptsize
\caption{Average Activity Values (Full table available online)\label{tab:table_average_activity}}

\begin{tabular}{lrrrrrrrc}
\hline
 Star &    S$_\mathrm{min}$ &     S$_\mathrm{max}$ &     S$_\mathrm{med}$ &      S$_\mathrm{STD}$ &  N$_\mathrm{obs}$ & B-V &  Rphk\_Noyes &  Age(Gyr) \\
\hline

     HD 10002 &  0.1585 &   0.1668 &  0.1598 &  0.00205 &    44 &  0.804 &     -5.041 &     6.45 \\
     HD 10008 &  0.3886 &   0.4526 &  0.4238 &  0.01347 &    37 &  0.768 &     -4.414 &     0.46 \\
    HD 100180 &  0.1615 &   0.1794 &  0.1679 &  0.00325 &    63 &  0.564 &     -4.916 &     4.12 \\
    HD 100623 &  0.1751 &   0.2165 &  0.1913 &  0.00930 &    82 &  0.763 &     -4.890 &     3.74 \\
    HD 101259 &  0.1392 &   0.1467 &  0.1443 &  0.00227 &    13 &  0.739 &     -5.118 &     8.37 \\
     HD 10145 &  0.1679 &   0.1769 &  0.1724 &  0.00150 &    27 &  0.660 &     -4.929 &     4.32 \\
    HD 101501 &  0.2769 &   0.3719 &  0.3135 &  0.02552 &    15 &  0.701 &     -4.526 &     0.99 \\
    HD 102158 &  0.1585 &   0.1601 &  0.1598 &  0.00050 &    12 &  0.557 &     -4.965 &     4.93 \\
    HD 103095 &  0.1910 &   0.2271 &  0.2087 &  0.00960 &    17 &  0.665 &     -4.774 &     2.44 \\
    HD 103432 &  0.2372 &   0.2716 &  0.2571 &  0.01037 &     9 &  0.645 &     -4.614 &     1.40 \\

\end{tabular}    
\end{table*}

\begin{table*}
\centering
\scriptsize
\caption{Chromospheric Time Series (Full Table (50k+) available online)\label{tab:table_time_series}}

\begin{tabular}{llrrllc}
\hline
      Star &    Filename &          BJD &  S-value &           SNR & Decker &  Seeing (arcsec) \\
\hline
    HD 185144 &     bj01.46 &  13237.736 &   0.2112 &           55 &     B5 &     1.0 \\
    HD 185144 &     bj01.47 &  13237.738 &   0.2120 &           34 &     B5 &     1.1 \\
    HD 185144 &     bj01.48 &  13237.739 &   0.2094 &           28 &     B5 &     1.1 \\
    HD 185144 &     bj01.49 &  13237.740 &   0.2118 &           30 &     B5 &     1.2 \\
    HD 185144 &     bj01.50 &  13237.740 &   0.2094 &           36 &     B5 &     1.0 \\
    HD 185144 &     bj01.51 &  13237.741 &   0.2112 &           43 &     B5 &     1.1 \\
    HD 185144 &     bj01.52 &  13237.742 &   0.2113 &           43 &     B5 &     1.0 \\
    HD 185144 &     bj01.53 &  13237.742 &   0.2114 &           48 &     B5 &     1.1 \\
    HD 185144 &     bj01.54 &  13237.743 &   0.2101 &           52 &     B5 &     1.0 \\
    HD 185144 &     bj01.55 &  13237.744 &   0.2115 &           49 &     B5 &     1.0 \\
\end{tabular}    
\end{table*}

\clearpage
\begin{longtable*}{lrrrrr}

\caption{Detected Stellar Cycles for 138 Stars \label{tab:table_cycles}} \\
\hline
Star &    Amplitude$_\mathrm{fit}$ &    Period$_\mathrm{fit}$ &   Threshold & Peak 1 & Peak 2 \\
\hline
\endfirsthead

\hline
Star &    Amplitude$_\mathrm{fit}$ &    Period$_\mathrm{fit}$ &   Threshold & Peak 1 & Peak 2 \\
\hline
\endhead

\hline
\multicolumn{6}{r}{{Continued on next page}} \\ 
\hline
\endfoot

\hline
\endlastfoot

  HD 100180 &  0.1704 &    3.33 &       1.76 &    1.01 &      0.36 \\
  HD 100623 &  0.2070 &    8.03 &       2.57 &    2.88 &      0.70 \\
  HD 103932 &  0.5085 &    9.00 &       1.36 &    5.49 &      0.47 \\
  HD 104304 &  0.1643 &   10.00 &       2.63 &    1.96 &      0.44 \\
   HD 10476 &  0.1977 &    9.11 &       1.39 &    0.86 &      0.26 \\
  HD 107148 &  0.1607 &    5.69 &       2.20 &    1.32 &      0.57 \\
  HD 109358 &  0.1674 &   13.50 &       1.87 &    1.09 &      0.36 \\
  HD 110315 &  0.3774 &   11.81 &       1.96 &    6.39 &      0.64 \\
  HD 111031 &  0.1496 &   13.72 &       1.29 &    0.53 &      0.28 \\
  HD 114783 &  0.2055 &    9.10 &       2.25 &    2.01 &      0.29 \\
  HD 116442 &  0.1686 &   18.09 &       2.49 &    1.84 &      0.21 \\
  HD 116443 &  0.1880 &   14.24 &       2.57 &    2.25 &      0.25 \\
  HD 122064 &  0.2863 &   12.81 &       2.91 &   11.62 &      1.06 \\
  HD 122120 &  0.6042 &   20.86 &       0.66 &    1.71 &      0.52 \\
  HD 125455 &  0.1946 &    9.87 &       3.39 &    5.65 &      0.57 \\
  HD 126053 &  0.1660 &   17.53 &       1.82 &    1.04 &      0.34 \\
  HD 126614 &  0.1469 &   16.15 &       2.24 &    1.19 &      0.45 \\
  HD 127334 &  0.1521 &   16.08 &       1.52 &    0.68 &      0.42 \\
  HD 130992 &  0.3386 &    3.11 &       0.71 &    0.69 &      0.42 \\
  HD 136713 &  0.3320 &    6.85 &       1.10 &    1.30 &      0.36 \\
  HD 139323 &  0.2467 &    8.84 &       2.28 &    2.57 &      0.23 \\
 HD 140538A &  0.2061 &    3.96 &       1.81 &    1.43 &      0.29 \\
   HD 14412 &  0.2032 &    6.15 &       2.51 &    2.68 &      0.63 \\
  HD 144287 &  0.1673 &   19.29 &       3.54 &    4.31 &      0.56 \\
  HD 145675 &  0.1883 &   11.20 &       3.49 &    4.24 &      0.21 \\
 HD 145958A &  0.1916 &    7.63 &       2.27 &    1.75 &      0.59 \\
 HD 145958B &  0.1868 &    8.30 &       1.79 &    1.15 &      0.75 \\
    HD 1461 &  0.1632 &   14.10 &       1.93 &    1.09 &      0.12 \\
  HD 146233 &  0.1735 &    6.27 &       3.30 &    3.87 &      0.97 \\
  HD 148467 &  0.7626 &    4.03 &       0.46 &    1.12 &      0.83 \\
  HD 149806 &  0.2244 &    6.71 &       2.08 &    1.99 &      0.95 \\
  HD 154088 &  0.1634 &   15.20 &       3.80 &    4.92 &      1.65 \\
  HD 154345 &  0.2320 &    6.95 &       2.82 &    5.58 &      0.40 \\
  HD 154363 &  0.5312 &    9.53 &       1.38 &    5.18 &      0.72 \\
  HD 155712 &  0.2296 &    9.33 &       3.15 &    5.12 &      0.85 \\
  HD 156279 &  0.1783 &   12.50 &       2.19 &    1.63 &      0.37 \\
  HD 156668 &  0.2462 &   11.38 &       1.95 &    2.37 &      0.19 \\
  HD 156985 &  0.3036 &    7.82 &       2.34 &    5.90 &      0.42 \\
  HD 158633 &  0.1780 &   11.83 &       2.13 &    1.51 &      0.90 \\
  HD 159062 &  0.1738 &   16.85 &       2.68 &    2.38 &      0.55 \\
   HD 16160 &  0.2371 &   12.43 &       2.31 &    3.40 &      0.48 \\
  HD 168009 &  0.1616 &   17.52 &       1.79 &    0.96 &      0.47 \\
  HD 170493 &  0.4691 &    8.75 &       1.26 &    3.33 &      0.98 \\
  HD 172051 &  0.1719 &    8.36 &       2.08 &    1.35 &      0.55 \\
   HD 17230 &  0.8034 &   15.20 &       0.24 &    0.53 &      0.13 \\
   HD 18143 &  0.1801 &   13.67 &       2.98 &    2.91 &      0.40 \\
  HD 182488 &  0.1704 &   11.54 &       2.79 &    2.34 &      0.35 \\
  HD 183263 &  0.1576 &    7.00 &       2.85 &    2.07 &      0.19 \\
  HD 185144 &  0.2179 &    5.93 &       1.94 &    1.78 &      0.20 \\
  HD 186408 &  0.1534 &   20.20 &       1.68 &    0.79 &      0.25 \\
   HD 18803 &  0.1894 &    5.04 &       2.20 &    1.75 &      0.31 \\
  HD 190406 &  0.1919 &   15.02 &       1.21 &    0.64 &      0.47 \\
  HD 191408 &  0.1978 &   16.77 &       2.56 &    2.56 &      0.76 \\
  HD 192310 &  0.2171 &   10.67 &       4.12 &   14.26 &      0.94 \\
   HD 19308 &  0.1627 &   22.28 &       3.11 &    2.86 &      0.40 \\
  HD 196761 &  0.1785 &   11.40 &       3.79 &    7.29 &      1.63 \\
  HD 197076 &  0.1886 &    5.31 &       3.17 &    4.27 &      2.33 \\
  HD 199305 &  1.6523 &    6.30 &       0.24 &    1.61 &      1.11 \\
  HD 201091 &  0.6256 &    7.17 &       0.68 &    1.75 &      0.38 \\
   HD 20165 &  0.2289 &    7.78 &       3.03 &    6.05 &      0.76 \\
  HD 202751 &  0.2474 &   12.49 &       3.03 &    6.29 &      0.52 \\
   HD 20619 &  0.2023 &    4.42 &       2.43 &    2.63 &      0.58 \\
  HD 208313 &  0.3012 &    5.96 &       1.80 &    2.57 &      0.68 \\
  HD 209458 &  0.1616 &    4.79 &       1.96 &    1.13 &      0.73 \\
  HD 210302 &  0.1609 &    5.85 &       1.65 &    0.83 &      0.55 \\
  HD 213042 &  0.4273 &    8.01 &       1.34 &    2.97 &      1.23 \\
  HD 215152 &  0.2642 &    8.04 &       1.71 &    2.03 &      0.62 \\
  HD 216259 &  0.1938 &   15.96 &       3.62 &    7.17 &      0.88 \\
  HD 216520 &  0.2015 &   19.19 &       2.25 &    2.11 &      0.48 \\
  HD 218566 &  0.2967 &    9.66 &       2.66 &    6.10 &      0.31 \\
  HD 218868 &  0.2133 &    4.84 &       2.43 &    2.78 &      0.40 \\
  HD 219134 &  0.2758 &   13.27 &       2.25 &    4.33 &      0.33 \\
  HD 219538 &  0.2516 &    7.09 &       1.88 &    2.24 &      1.26 \\
 HD 219834B &  0.2059 &    9.70 &       3.61 &    6.77 &      0.77 \\
  HD 220339 &  0.2675 &    5.87 &       2.17 &    3.03 &      1.25 \\
  HD 221354 &  0.1599 &   18.84 &       1.51 &    0.71 &      0.19 \\
  HD 224619 &  0.1702 &   18.06 &       2.57 &    1.99 &      1.04 \\
  HD 239960 &  1.4673 &   15.78 &       0.32 &    0.88 &      0.49 \\
   HD 24496 &  0.2008 &    5.86 &       2.87 &    3.27 &      0.58 \\
   HD 25329 &  0.1952 &    6.01 &       2.36 &    2.16 &      0.33 \\
   HD 25665 &  0.3042 &    6.52 &       1.68 &    2.47 &      0.53 \\
   HD 26151 &  0.2061 &   16.35 &       3.24 &    5.25 &      0.40 \\
   HD 26161 &  0.1531 &   20.12 &       1.79 &    0.87 &      0.37 \\
   HD 26965 &  0.2060 &    9.11 &       2.13 &    1.92 &      0.42 \\
   HD 28005 &  0.1633 &   13.79 &       2.80 &    2.28 &      0.68 \\
   HD 28946 &  0.2443 &    5.90 &       2.38 &    3.29 &      1.49 \\
   HD 29883 &  0.1958 &   14.00 &       3.48 &    6.49 &      0.42 \\
   HD 31253 &  0.1526 &   13.08 &       1.48 &    0.66 &      0.43 \\
   HD 32147 &  0.2933 &    9.53 &       0.93 &    0.74 &      0.26 \\
   HD 34445 &  0.1665 &   23.01 &       1.69 &    0.91 &      0.23 \\
   HD 36003 &  0.4252 &   10.98 &       1.44 &    3.64 &      0.72 \\
   HD 36395 &  2.0485 &    2.30 &       0.21 &    1.63 &      1.22 \\
    HD 3651 &  0.1788 &   10.01 &       3.14 &    3.38 &      0.35 \\
   HD 37008 &  0.1821 &   21.27 &       1.37 &    0.76 &      0.27 \\
    HD 3765 &  0.2186 &   12.80 &       3.18 &    5.96 &      0.31 \\
   HD 38230 &  0.1642 &   23.15 &       2.32 &    1.54 &      0.81 \\
   HD 38529 &  0.1734 &    5.46 &       2.63 &    2.02 &      0.54 \\
    HD 4256 &  0.2639 &    9.66 &       1.73 &    1.93 &      0.32 \\
   HD 42618 &  0.1639 &   10.16 &       2.83 &    2.30 &      0.19 \\
   HD 43947 &  0.1586 &   12.88 &       2.12 &    1.26 &      0.52 \\
    HD 4628 &  0.2173 &    7.91 &       2.73 &    3.46 &      0.81 \\
    HD 4747 &  0.2676 &    4.90 &       1.43 &    1.41 &      0.65 \\
    HD 4915 &  0.2046 &    4.86 &       1.72 &    1.19 &      0.42 \\
   HD 49674 &  0.1977 &    4.05 &       1.12 &    0.62 &      0.36 \\
   HD 50499 &  0.1516 &    3.69 &       2.51 &    1.56 &      0.76 \\
   HD 51419 &  0.1948 &   23.43 &       2.09 &    1.67 &      0.23 \\
   HD 51866 &  0.3395 &    7.17 &       0.98 &    1.10 &      0.30 \\
   HD 52711 &  0.1602 &   13.96 &       2.91 &    2.42 &      0.49 \\
   HD 62613 &  0.2089 &    5.75 &       3.27 &    5.28 &      0.90 \\
   HD 65277 &  0.2543 &   12.19 &       2.66 &    6.53 &      0.51 \\
   HD 68988 &  0.1630 &    5.29 &       2.67 &    1.99 &      0.62 \\
   HD 69830 &  0.1734 &   10.83 &       1.84 &    1.12 &      0.60 \\
   HD 72673 &  0.1866 &   10.47 &       4.03 &    9.65 &      1.67 \\
   HD 73667 &  0.1717 &   19.00 &       2.33 &    1.64 &      0.59 \\
   HD 74156 &  0.1476 &   16.01 &       2.05 &    1.05 &      0.19 \\
   HD 75732 &  0.1973 &   11.02 &       3.79 &    6.74 &      0.27 \\
    HD 7924 &  0.2285 &    7.49 &       2.40 &    3.15 &      0.27 \\
   HD 80606 &  0.1566 &   16.27 &       1.96 &    1.06 &      0.34 \\
   HD 82943 &  0.1716 &    2.96 &       1.56 &    0.84 &      0.28 \\
    HD 8389 &  0.2158 &   10.67 &       3.24 &    4.31 &      0.42 \\
   HD 84035 &  0.5291 &    8.03 &       0.95 &    2.16 &      0.73 \\
   HD 87359 &  0.2047 &    4.40 &       1.91 &    1.64 &      0.82 \\
   HD 87883 &  0.2780 &    7.83 &       1.85 &    2.93 &      0.52 \\
   HD 89269 &  0.1698 &   16.95 &       2.84 &    2.63 &      0.37 \\
   HD 90875 &  0.7309 &   18.71 &       0.84 &    2.38 &      0.62 \\
    HD 9562 &  0.1509 &   15.39 &       1.57 &    0.71 &      0.29 \\
   HD 97658 &  0.2074 &    9.01 &       3.64 &    8.63 &      0.30 \\
   HD 98281 &  0.1801 &   15.51 &       2.53 &    2.33 &      0.94 \\
   HD 99491 &  0.2128 &    6.15 &       2.61 &    3.26 &      0.33 \\
   HD 99492 &  0.2820 &    9.81 &       1.97 &    2.81 &      0.19 \\
    HD 9986 &  0.1794 &    5.03 &       1.87 &    1.23 &      0.68 \\
      GL239 &  1.0562 &   15.24 &       0.38 &    1.39 &      0.86 \\
      GL273 &  0.8215 &    2.94 &       0.30 &    0.66 &      0.33 \\
      GL699 &  0.8470 &    8.48 &       0.27 &    0.50 &      0.31 \\
   HIP19165 &  0.7498 &    5.68 &       0.43 &    0.94 &      0.69 \\
   HIP41689 &  0.9481 &    3.03 &       0.23 &    0.56 &      0.38 \\
   HIP74995 &  0.5778 &    3.80 &       0.38 &    0.58 &      0.17 \\
    S130811 &  1.1601 &   13.58 &       0.45 &    2.27 &      0.43 \\

\end{longtable*}




 




\clearpage
\bibliography{main}{}
\bibliographystyle{aasjournal}

\end{document}